\pdfoutput=1 

\documentclass[
     a4paper,
     ]{article}

\usepackage{IEK10} 
\usepackage{natbib}

\newcommand{\mytitle}{Predicting the Temperature Dependence of Surfactant CMCs Using \\Graph Neural Networks}

\newcommand{\affil}{
  \begin{itemize}[leftmargin=3mm, itemsep=0mm]
        \item[$^a$]BASF Personal Care and Nutrition GmbH, Henkelstrasse 67, 40589 Duesseldorf, Germany %
        \item[$^b$]RWTH Aachen University, Process Systems Engineering (AVT.SVT), Aachen, Germany %
		\item[$^c$]Forschungszentrum J\"ulich GmbH, Institute for Energy and Climate Research IEK-10: Energy Systems Engineering, J\"ulich, Germany%
		\item[$^d$]JARA-ENERGY, Aachen, Germany
  \end{itemize}
}

\def\firstAuthor{Christoforos Brozos}
\newcommand{\myauthor}{
	Christoforos Brozos$^{a,b}$, 
	Jan G. Rittig$^b$,
	Sandip Bhattacharya$^a$,
    Elie Akanny$^a$,
    Christina Kohlmann$^a$,
	Alexander Mitsos$^{d,b,c,*}$ %
}
\newcommand{\correspondingauthor}{
    Corresponding author, E-mail: amitsos@alum.mit.edu
}
\author{\myauthor}

\usepackage[hyphens]{url}
\usepackage[
  colorlinks,
  linkcolor=blue,
  citecolor=blue,
  urlcolor=blue,
  pdftitle={\mytitle},
  pdfauthor={\firstAuthor}
]{hyperref}
\usepackage[capitalise, nameinlink]{cleveref}
\usepackage[export]{adjustbox}
\crefname{table}{Tab.}{Tab.}

\begin{document}

	\thispagestyle{firststyle}
	
	\begin{center}
		\begin{large}
			\textbf{\mytitle}
		\end{large} \\
		\vspace{0.1cm}
		\myauthor
	\end{center}
	
	\vspace{-0.4cm}
	
	\begin{footnotesize}
		\affil
	\end{footnotesize}
	
	\vspace{-0.3cm}

	\section*{Abstract}
	
	The critical micelle concentration (CMC) of surfactant molecules is an essential property for surfactant applications in industry. Recently, classical QSPR and Graph Neural Networks (GNNs), a deep learning technique, have been successfully applied to predict the CMC of surfactants at room temperature. However, these models have not yet considered the temperature dependency of the CMC, which is highly relevant for practical applications. We herein develop a GNN model for temperature-dependent CMC prediction of surfactants. We collect about 1400 data points from public sources for all surfactant classes, i.e., ionic, nonionic, and zwitterionic, at multiple temperatures. We test the predictive quality of the model for following scenarios: i) when CMC data for surfactants are present in the training of the model in at least one different temperature, and ii) CMC data for surfactants are not present in the training, i.e., generalizing to unseen surfactants. In both test scenarios, our model exhibits a high predictive performance of R$^2 \geq $ 0.94 on test data. We also find that the model performance varies by surfactant class. Finally, we evaluate the model for sugar-based surfactants with complex molecular structures, as these represent a more sustainable alternative to synthetic surfactants and are therefore of great interest for future applications in the personal and home care industries.
	
	\vspace{0.1cm}



\section{Introduction}\label{sec:intro}

\noindent Surfactants are used in a wide variety of industries such as cosmetics, food additives, detergents, oil recovery, and pharmaceuticals due to their unique properties~\citep{Vieira2021,Shaban2020,Nitschke2007,2005,Adu2020,Szuets2012,Massarweh2020}. 
Surfactants are amhiphilic molecules containing hydrophilic (head) and hydrophobic (tail) parts. A surfactant property of major interest is the critical micelle concentration (CMC), which is the concentration above which self-aggregation into micelles takes place and is highly relevant in many applications~\citep{Su2020,Thompson2023,Ghezzi2021,Kumar2019}. The CMC is accompanied by sharp changes in the bulk solution properties~\citep{Rosen2012,MyersAugust2020}. 

The influence of temperature on CMC is of great importance in surfactant applications, such as washing and oil recovery. Specifically, the CMC is influenced by multiple parameters that are related to the surfactant itself, such as the structure of the tail and nature of the head group, and parameters related to the solute environment, such as pH, temperature, and the presence of electrolytes~\citep{Rosen2012,MyersAugust2020,Katritzky2008,Thiruvengadam2020}. The temperature is of particular interest, since no model applicable to all surfactant classes is available in the literature. Many researchers have studied the temperature effect on the CMC, which in general has been described as complex and varies for each surfactant class~\citep{Rosen2012,MyersAugust2020}. For nonionics, authors have reported that CMC decreases monotonically as the temperature increases~\citep{Lindman2016}, but in general C$_{i}$E$_{j}$ (polyethylene oxide) type of surfactants exhibit a minimum CMC at temperature around 50$^\circ$C~\citep{Rosen2012,MyersAugust2020,Kroll2022,Chen1998}. In contrast, MEGA (N-alkanoyl-N-methylglucamide) surfactants show a U-shaped relationship~\citep{Prasad2006,Okawauchi1987}. Ionic surfactants normally exhibit a U-shaped relationship too, with their minimum CMC value being between 20-30$^\circ$C~\citep{Rosen2012,Perger2007,Fu2019}. It has been recently shown by \cite{Brinatti2014} that increasing the temperature causes a decrease in the CMC of some zwitterionics and others to exhibit a minimum CMC. The latter is also shown in the work of Mukerjee and Mysels~\citep{Mukerjee1971}. All the above are experimental observations for individual surfactants, but no model for explicit CMC predictions at multiple temperatures of a wide spectrum of surfactants has been derived. This leads to limitations in predicting the temperature-dependent CMC of new surfactant molecules and highlights the need to consider the surfactant structure in the development of predictive models. 

Both bio-based surfactants and biosurfactants are of particular interest in personal and home care industries since they enable a transition from fossil to renewable feedstocks~\citep{Farias2021,Jahan2020}. In the literature, biosurfactants are defined as those produced by biotechnology while bio-based surfactants are surfactants comprised of sustainable 100\% natural-based feedstocks~\citep{DRAKONTIS202077}. Both bio-based surfactants and biosurfactants contain sugar-based groups, hence are also referred to as sugar-based surfactants. The general structure comprises a polar carbohydrate, like glucose, as the hydrophilic part and an alkyl chain as the hydrophobic part~\citep{ruiz2008sugar}. \cite{Gaudin2019} recently discussed the impact of this chemical structure on the sugar-based surfactant properties. Specifically, they state that the CMC can be affected by the stereochemistry and the anomeric configuration of the sugar head. Although sugar-based surfactants belong to nonionics, the effect of temperature on their solution properties differs from the ethoxylated ones~\citep{MolinaBolivar2004}. Researches have reported a decrease in CMC with the increase of temperature~\citep{MolinaBolivar2004} and a U-shaped relationship in other cases~\citep{Paula1995,Angarten2014,Castro2018}, thus showing again a complex relationship between surfactant structure, CMC and temperature.

The importance of CMC as a key characteristic surfactant property has driven many researchers to investigate the relationship between surfactant structure and CMC~\citep{Rosen2012}. Empirical equations have been derived from experimental data to describe this relationship, but they are only applicable to specific systems~\citep{Rosen2012,Klevens1953}. Recently, numerous quantitative structure-property relationship (QSPR) type models were developed for CMC prediction~\citep{Katritzky2008,Gaudin2016,Hu2010,Mattei2013,Wang2018,Abooali2023,Seddon2022}. In QSPR models, molecular descriptors are estimated from the molecular structure and mapped to the target property. One limitation is that they typically require manual selection and identification of effective molecular descriptors.

An alternative end-to-end deep learning approach, called graph neural networks (GNNs), has been successfully applied to numerous molecular property prediction tasks~\citep{Schweidtmann2020,Rittig2022,Gilmer2017,Rittig2023,SanchezMedina2023,SanchezMedina2022,Yang2019}. GNNs are applied directly to the molecular graph and extract the necessary structural information which they later use to predict the target property, thereby providing an end-to-end learning framework~\citep{hamilton2020graph}. Due to their broad success and adoption, GNNs have been effectively applied to predict the CMC and surface excess concentration ($\Gamma$$_{m}$) of surfactant monomers~\citep{Qin2021,brozos2024graph,Moriarty2023}. For both approaches, i.e., QSPR and GNNs, the temperature-dependency of CMC is rarely studied. In fact, the effect of the temperature on the CMC has only been modeled in one recently published QSPR model, which is however limited to anionic surfactants~\citep{Abooali2023}. We therefore aim at investigating the prediction of the temperature effect on CMC for all common surfactant types namely ionic, nonionic, and zwitterionic classes.  

In the present work, we develop a GNN model for predicting the temperature-dependent CMC of all surfactant classes. To train the models, we extend our database of CMC values at room temperature for different surfactant classes from our previous work by adding CMC measurements at various from publicly available sources~\citep{brozos2024graph}. The extended database contains around 1400 measurements from 492 unique surfactant molecules. We then develop a GNN model that learns to extract structural information from the surfactant molecules in a fixed-size vector, known as the molecular fingerprint, in which we concatenate the experimental temperature. The updated vector is then mapped to the measured CMC of the surfactant. We incorporate stereochemistry and anomeric configuration information into the molecular graph representation. We train and fine-tune our model on the newly constructed database. We then test our model to predict the temperature-dependent CMC of i) surfactant structures seen during training but at different temperatures, and ii) completely new surfactants, i.e., not seen during training. Furthermore, we particularly explore the temperature dependency of the CMC of sugar-based surfactants. We show that in both test cases, the GNN model provides accurate predictions.

The remainder of this work is constructed as follows: First a description of the collected data set and of the test sets is provided in Section~\ref{sec:data_sets}. In Section~\ref{sec:methods}, we present the concept of the GNN for predicting temperature-dependent CMCs. Subsequently, the predictive quality of the developed GNN model is presented, discussed, and compared with previous works in Section~\ref{sec:res_disc}. Finally, we conclude this contribution in Section~\ref{sec:conclusion}. The test data used for model evaluation is publicly available in our~\href{https://github.com/brozosc/Predicting-the-Temperature-Dependence-of-Surfactant-CMCs-using-Graph-Neural-Networks}{GitHub repository} and the model architecture is presented in Section~\ref{sec:methods}. The training data set and the trained models remain property of BASF and could be made available upon request.

\section{Data set}\label{sec:data_sets}

\noindent 

\subsection{{Data set overview}}
We collected a data set of temperature-dependent CMC values for a variety of surfactants. Particularly, we extended the assembled CMC data set from our previous work~\citep{brozos2024graph}, to include CMC information at multiple temperatures for each molecule, when such data was available. In the data collection process, we prioritized CMC values measured through tensiometry and we excluded any duplicates. For those surfactants for which we did not find tensiometry data, we considered CMC values determined by conductometry, light scattering, refractive index and calorimetry. Each data point includes the CMC, the temperature, and the isomeric SMILES string of the surfactant, allowing to distinguish between different anomers, e.g., octyl-$\alpha$/$\beta$-D-glycoside, and chiral centers in the sugar head, e.g., glucoside with galactoside~\citep{Weininger1988}. In total our new data set consists of 1,377 data points, with 492 unique surfactant structures of which 227 structures were measured at least at two different temperatures. The minimum temperature of our data is 0$^\circ$C and the maximum is 90$^\circ$C. The temperature distribution of the data set is illustrated in Figure~\ref{fig:temperature_distribution}. A detailed statistical analysis of the CMC data can be found in the Appendix A, cf. Figure~\ref{fig:apendix_cmc_distribution}.

\begin{figure}
    \centering
    \includegraphics[height = 10 cm, width = 13 cm]{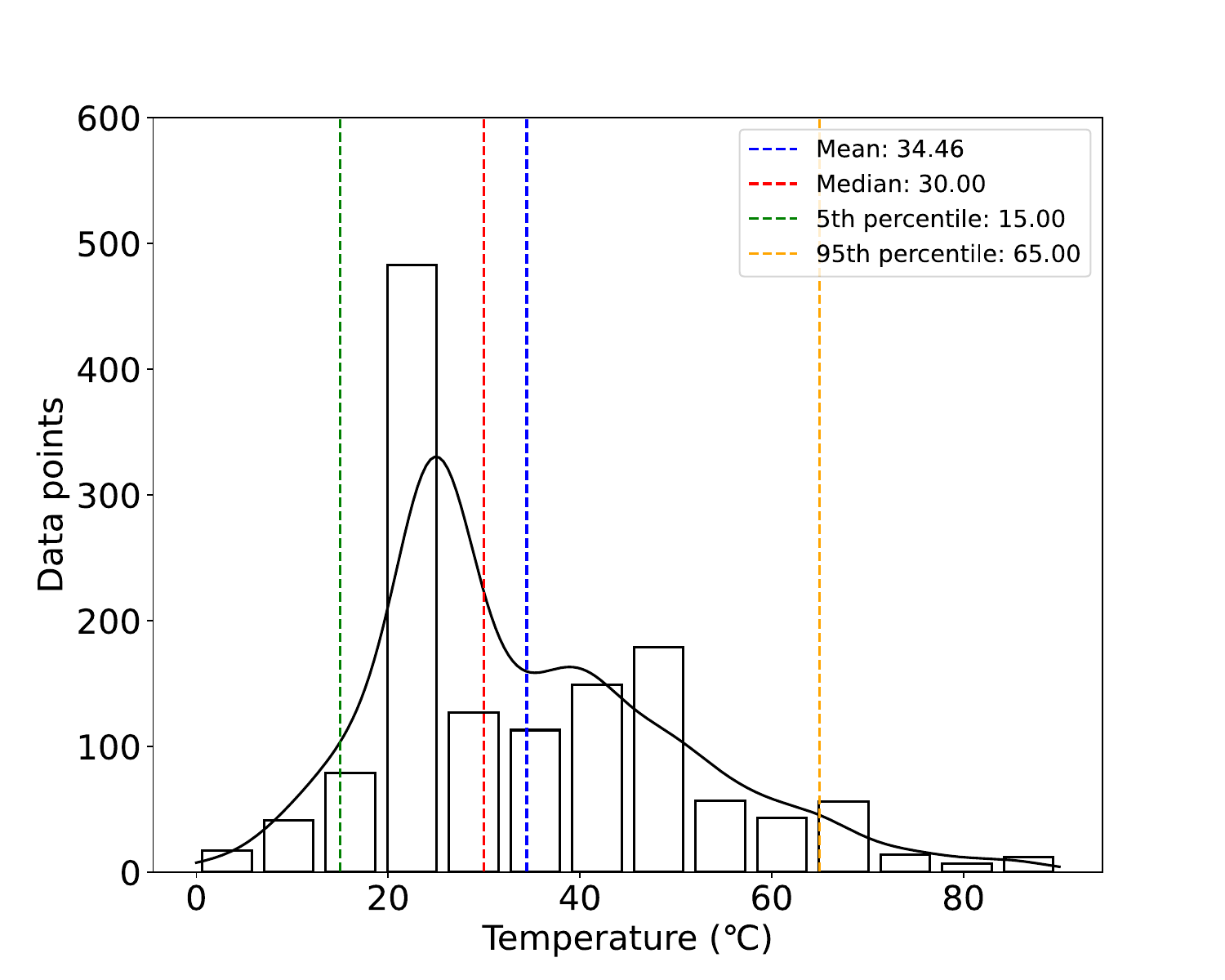}
    \caption{Distribution of the CMC data with the number of data points over the temperature range of T$=10-90$ for a bin size of 10$^\circ$C.}
    \label{fig:temperature_distribution}
\end{figure}

\begin{figure}
\centering
\subfloat[S1]{\includegraphics[height = 5.5cm, width = 8cm]{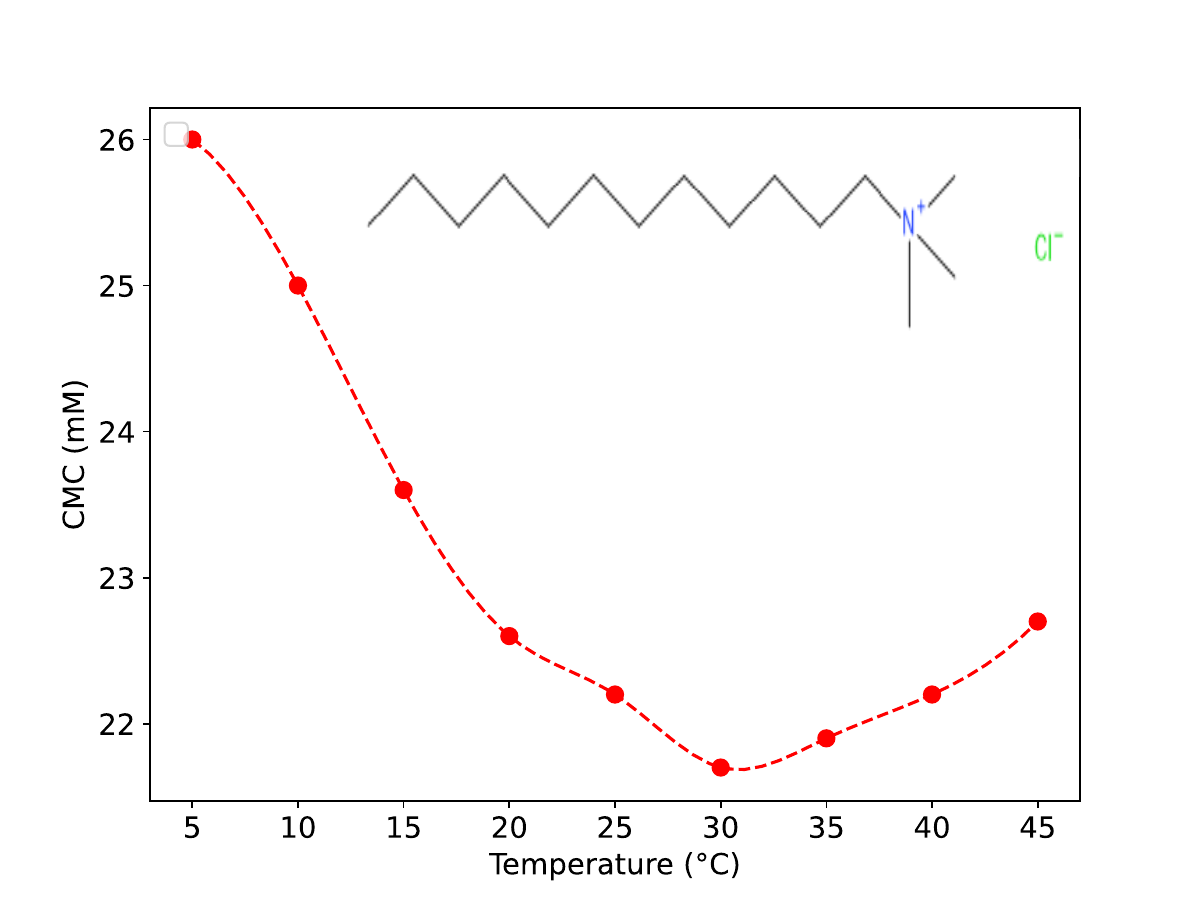}}
\subfloat[S2]{\includegraphics[height = 5.5cm, width = 8cm]{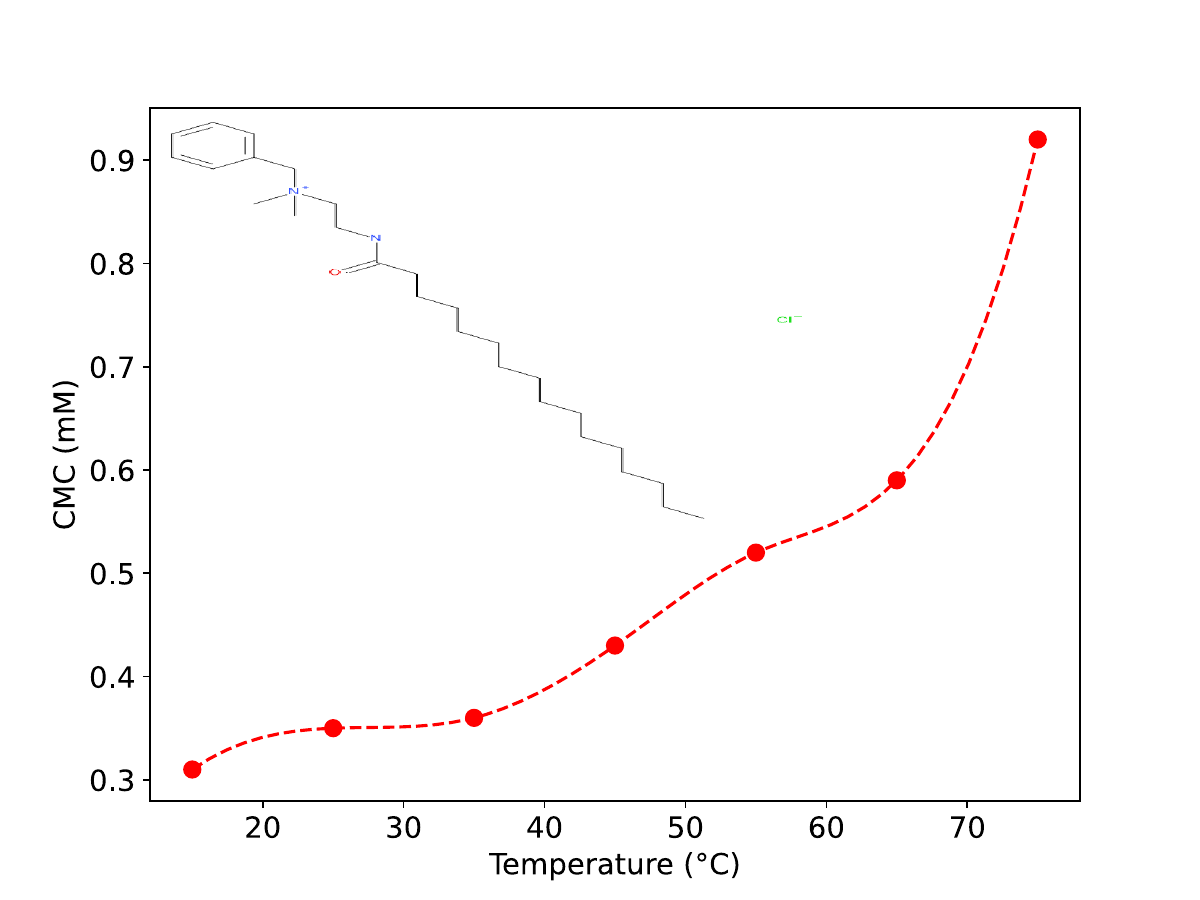}}
\hfill
\subfloat[S3]{\includegraphics[height = 5.5cm, width = 8cm]{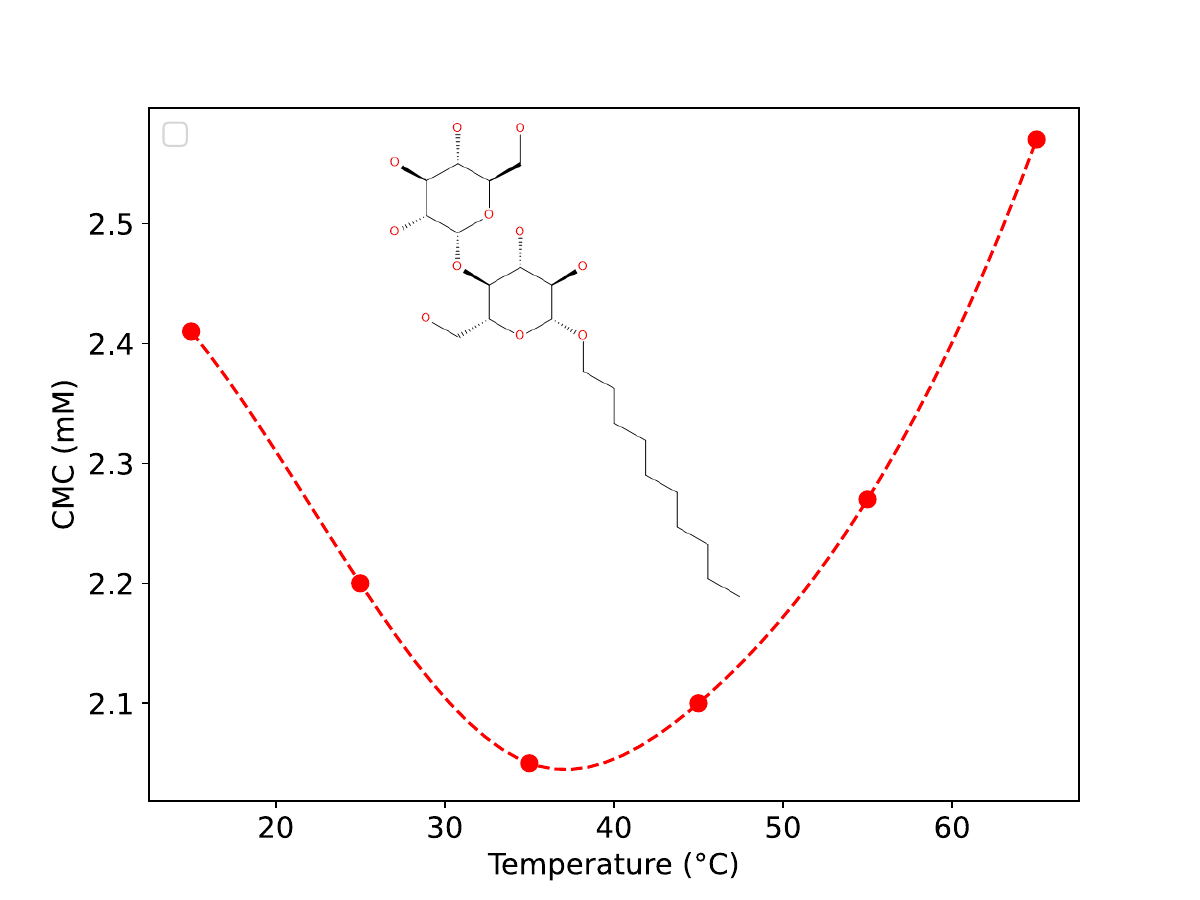}}
\subfloat[S4]{\includegraphics[height = 5.5cm, width = 8cm]{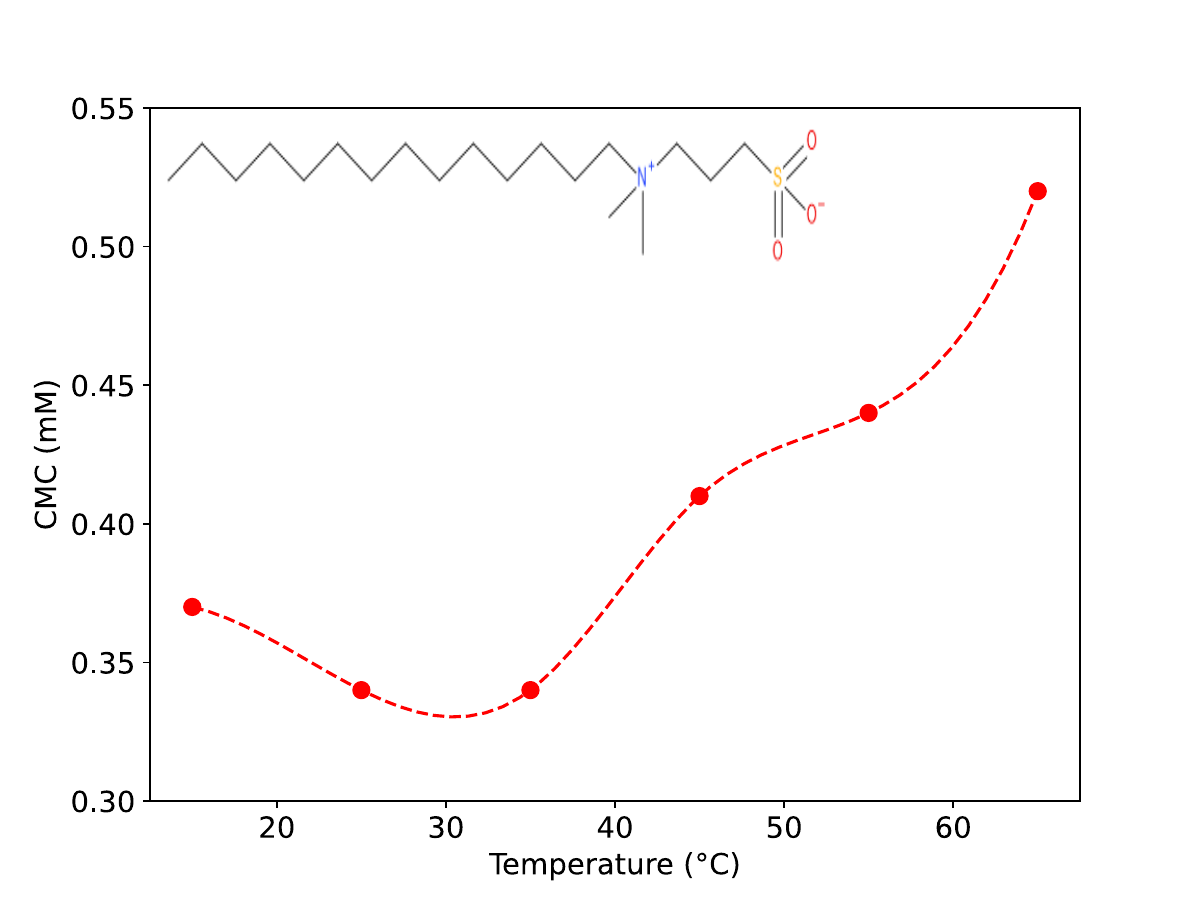}}
\hfill
\subfloat[S5]{\includegraphics[height = 5.5cm, width = 8cm]{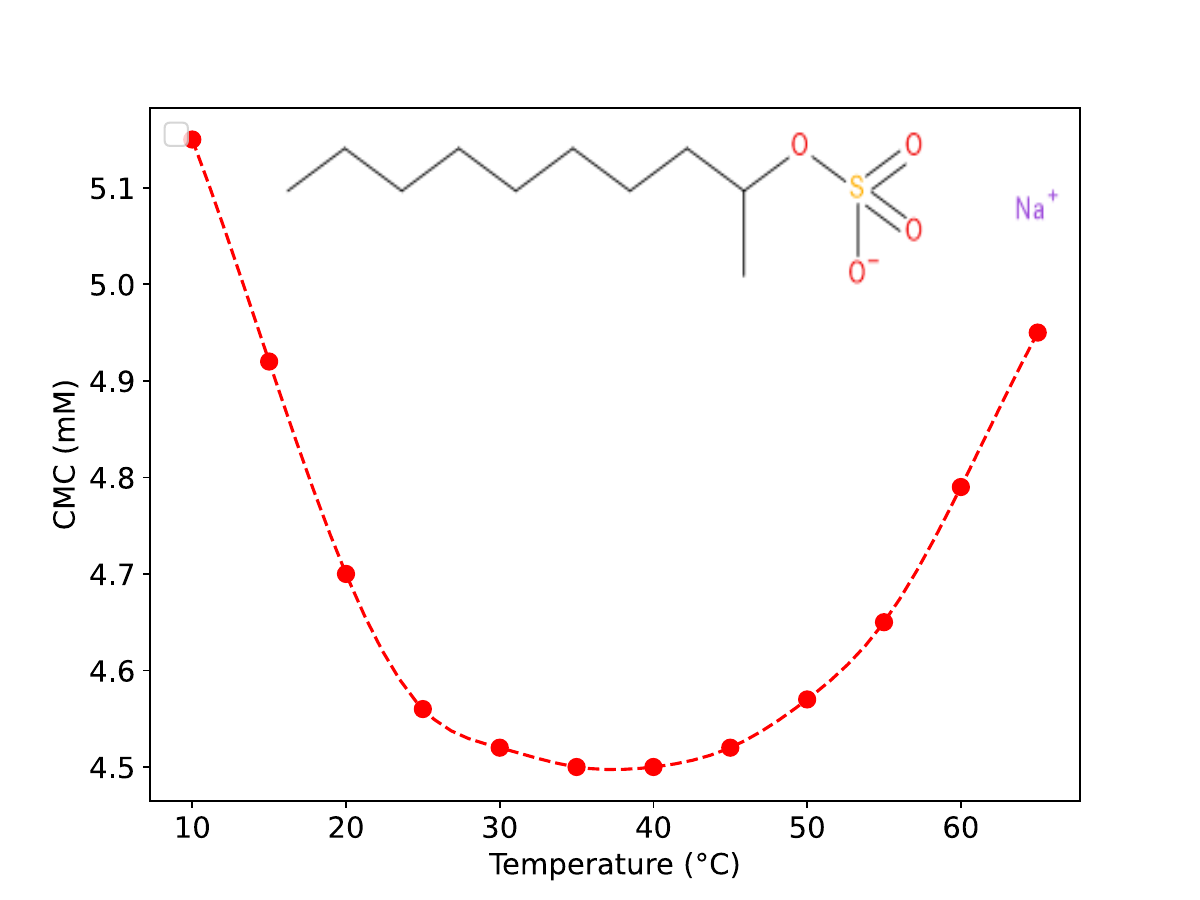}}
\subfloat[S6]{\includegraphics[height = 5.5cm, width = 8cm]{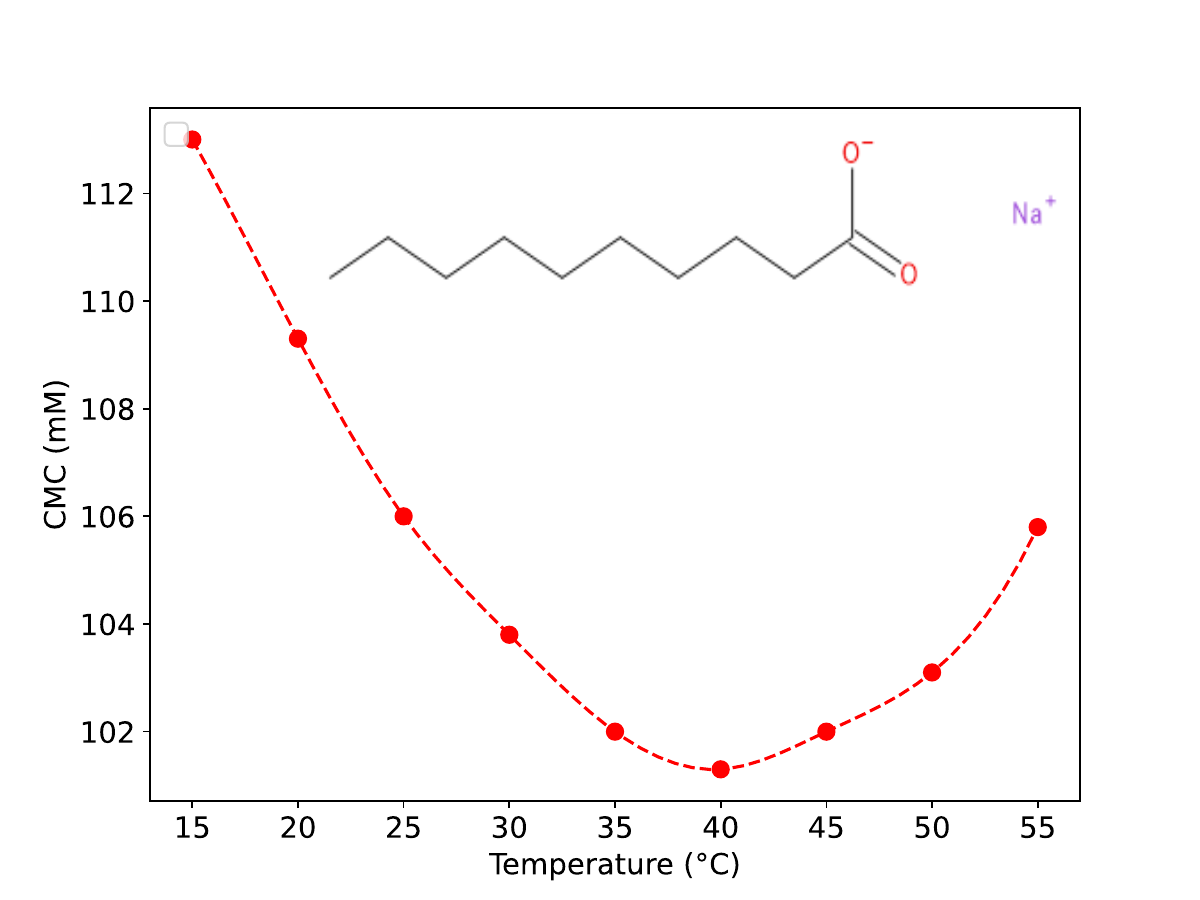}}
\caption{Experimental values of CMC (in mM) at different temperatures for six surfactants present in our database are tabulated: (S1) dodecyltrimethylammonium chloride (DTAC)~\citep{Perger2007}, (S2) benzyl (3-hexadecanoylaminoethyl)dimethylammonium chloride (C15AEtBzMe2Cl)~\citep{Galgano2010}, (S3) Decyl diglucoside~\citep{Brinatti2014}, (S4) Sulfobetaine 14~\citep{Cheng2012}, (S5) Sodium decyl 2 sulfate~\citep{Mukerjee1971} and (S6) Sodium decanoate~\citep{GonzalezPerez2005}. Each of these surfactants shows a different temperature dependence.}
\label{fig:descriptive_examples}
\end{figure}

\par The different relationships in each surfactant classes between temperature and CMC discussed in Section~\ref{sec:intro} are also present in our data set. In Figure~\ref{fig:descriptive_examples}, six surfactant examples are illustrated. Specifically, the three ionic surfactants, namely S1, S5, and S6 exhibit a U-shaped relationship, as often is the case for ionic surfactants. The minimum CMC is between 30 and 40$^\circ$C. An exception to the U-shaped relationship is S2, an ionic surfactant too, as an increase in temperature causes the CMC to increase too. Finally, the sugar-based nonionic surfactant S3 and the zwitterionic surfactant S4 exhibit a U-shaped relationship.

\begin{table}
\centering
\caption{A description of the nature of the data set used in this paper. The number of surfactants per class in the complete data set and the test sets are listed.}
\label{tab:datasets} 
\begin{tabular}{l | c |  c c}
	& \textbf{Full data set} & \multicolumn{2}{ c }{\textbf{Test sets}} \\
    &  & \textbf{Different temperature} & \textbf{Distinct surfactant} \\
    \hline 
    Anionics & 576 & 100 & 89 \\
		
	Nonionics & 422 & 62 & 68\\
		
	Cationics & 291 & 49 & 50\\
		
	Zwitterionics & 88 & 16 & 11\\
	\hline
	\textbf{Total data points} & 1,377 & 227 & 218 \\
\end{tabular}
\end{table}

\subsection{Data splits}\label{sec:split_types}
We implement two types of data set splitting: i) \textbf{\emph{different temperature}} -- for testing the prediction accuracy of our model at new temperatures; ii) \textbf{\emph{distinct surfactant}} -- for testing the ability of our model to generalize to new, unseen surfactant structures, hence structures not included in the model training at all. We note that the two test sets have a similar surfactant class distribution in Table~\ref{tab:datasets}, which was not enforced.
\par For selecting the \emph{different temperature} test set, we first identify all unique surfactant molecules with CMCs in at least two different temperatures, and then randomly select one data point for each of these molecules to be in the test set. For example, if for a given surfactant there exist 3 measurements at 3 different (distinct) temperatures in the full data set, one will be randomly assigned to the test set and the other two remain in the training set. A total of 227 molecules at various temperatures are selected for testing, which accounts for about 16\% of the whole data set size.

The \emph{distinct surfactant} test set aims to evaluate the models' predictive performance for completely unseen surfactant molecules at various temperatures. For consistency in the performance comparison, we choose similar test set size, namely 218 data points. We randomly select molecules and add all corresponding data points to the test, such that the temperature-dependent CMC values of these molecules remain completely unseen during training. In this final test set, about 70\% of the data points were molecules measured at different/multiple temperatures. For these molecules, we include all available CMC values measured at different temperatures in the test set. The remaining 30\% of the test set are molecules measured only at one temperature. In total, 50 distinct surfactant structures are included in the test set.

\section{Methods}\label{sec:methods}
\noindent
In the following sections, we first present the structure of our GNN model and describe ensemble learning (Section~\ref{sec:graph_neural}). Then, we summarize the hyperparameter selection and training settings of our model (Section~\ref{sec:hyperparameter_selection}). 

\subsection{Graph Neural Networks and ensemble learning}\label{sec:graph_neural}
\begin{figure}[htbp]
	\centering
	\includegraphics[height = 7cm, width = 14 cm]{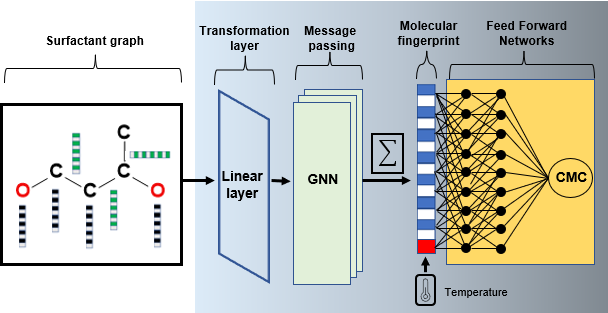}
	\caption{Schematic representation of the developed graph neural network for predicting temperature-dependent CMC values of surfactant monomers.}
	\label{fig:model_overview}
\end{figure}

\noindent
We adapt our GNN model for predicting CMC values of surfactant monomers from our previous work~\citep{brozos2024graph} to include the temperature dependency. An overview of the model framework is illustrated in Figure~\ref{fig:model_overview}. The GNN takes as input a surfactant molecule, represented as an undirected molecular graph, where atoms correspond to nodes (vertices) and bonds to edges. Each node and each edge is assigned a feature vector, where chemical information about the corresponding atom/bond is stored. The node and edge features used in the current work are presented in Tables~\ref{tab:appendix_node_features} and~\ref{tab:appendix_edge_features} respectively. GNNs operate directly on the node and edge features of the undirected molecular graph~\citep{Gilmer2017}. During graph convolutions, the node features, denoted as hidden states, are updated with structural information from their neighborhood. Then, a pooling step is applied. That is, the updated hidden state of all nodes within the graph, which now contain information about the respective node itself and the neighbor nodes, are combined, e.g., by applying the sum operator, into a single unique vector representation, known as the molecular fingerprint. To include the temperature, we first normalized it between a minimum of 0 and a maximum of 10 and we concatenated the molecular fingerprint with it. Finally, the updated molecular fingerprint is used to perform the downstream property prediction task, herein predicting the CMC, in a standard multilayer perceptron (MLP). For a detailed overview of GNNs, including different types of graph convolutions and pooling operators we refer to works on graph convolutions layers (cf.~\citep{hamilton2018inductive,velickovic2018graph,xu2019powerful,simonovsky2017dynamic}) and pooling operators (cf.~\citep{cangea2018sparse,vinyals2016order, Schweidtmann_PoolingGNN.2022,Vu2023,Buterez2023}), also see reviews in~\citep{Zhou2020,9046288}.

In our previous work, we considered stereochemistry information on the edge feature vector~\citep{brozos2024graph}. As both chirality and stereochemistry of a surfactant molecule impact the CMC, we herein additionally use chirality information on the atom feature vector too. We note that more sophisticated approaches exist in the literature~\citep{Adams2021Learning3R} and that using 3D information could be beneficial but requires to determine or calculate 3D coordinates and conformers which is computationally expensive. In our work, we aim to investigate whether the GNN model that learns from molecular graphs with chirality information can perform accurate predictions of CMC in different anomers and isomers too.

We further apply ensemble learning. Ensemble learning is a common technique in machine learning to reduce the noise of randomly chosen training and validation sets. Multiple models are trained, i.e., on different splits of training and validation sets, i.e., non-test data, and their predictions are averaged, thus leading to more robust and generalized predictions~\citep{Breiman1996,Dietterich2000,Ganaie2022}. We herein train 40 different models on both split types mentioned in Section~\ref{sec:split_types}, and then we average out the predictions to report the prediction accuracy of the ensemble of GNNs.

\subsection{Implementation and hyperparameter}\label{sec:hyperparameter_selection}
\noindent
To determine the optimal hyperparameter values, we train our GNN model on 40 different seeded validation sets, similar to our previous works~\citep{Schweidtmann2020,Rittig2022,Rittig2022a,brozos2024graph}. The size of the validation set is kept constant at 200 molecules, which represents about 14\% of the whole data set size and thus a general training-validation-test split of 70:14:16. We use the \emph{different temperature} test set for our hyperparameter tuning and we select the ones leading to the minimum root mean squared error (RMSE) on the validation set. We scale the CMC ($\mu$M) values using a (based 10) logarithmic scale. We use a grid search to investigate the hyperparameters of the GNN model described in Table~\ref{tab:appendix_model_hyperparameters}. Note that the first layer of the MLP has a size of 129 neurons because we are concatenating the normalized temperature on the molecular fingerprint

We represent each surfactant molecule with an isomeric SMILES string~\citep{Weininger1988}. We use RDKit (\emph{version 2022.3.5}), an open-source toolkit for cheminformatics to generate the attributed molecular graph for each surfactant. For the graph convolutions, the GINE-operator~\citep{xu2019powerful,hu2020strategies} as implemented in PyTorch Geometric (PyG)~\citep{Fey2019} is applied with sum being the pooling layer of choice. We use sum pooling as it has been shown to be beneficial for molecular-dependent prediction tasks~\citep{Schweidtmann_PoolingGNN.2022}.

\section{Results and discussion}\label{sec:res_disc}
\noindent 
In this section, the predictive performance of our GNNs models is evaluated (Section~\ref{sec:pred_new_temperatures}). Then, we compare our GNN model with predictive CMC models from previous works (Section~\ref{sec:prev_work}). We further analyze the predictive performance for different temperature ranges (Section~\ref{sec:temp_accuracy}) and investigate the influence of the surfactant classes on the CMC predictions (Section~\ref{sec:performance_class}). Finally, the results on selected sugar-based surfactants are presented, which are interesting due to their sustainability aspects (Section~\ref{sec:sugar_results}).

\subsection{Predictive performance on different temperature and distinct surfactant test sets}\label{sec:pred_new_temperatures}
\noindent
We first evaluate the model performance for two different test scenarios (cf. Section~\ref{sec:split_types}): the \emph{different temperature}, i.e., predicting the CMC of surfactants that have been included in training but at different temperatures, and the \emph{distinct surfactant}, i.e., predicting the CMC at different temperatures of surfactants that have not been used for training at all.
Table~\ref{tab:accuracies} shows the mean absolute error (MAE), the RMSE, and the mean absolute percentage error (MAPE) on the validation and the test set for both scenarios. 
Here, we report the performance of the ensemble of GNNs since we found ensemble learning to be beneficial in our previous work~\citep{Schweidtmann2020,brozos2024graph}. All reported error metrics refer to the log CMC.

\begin{table}[b]
	\centering
	\caption{Summary of GNN ensemble accuracy on the two test set splits for predicting the temperature-dependent CMC on logarithmic scale. The ensembles consist of 40 GNNs. The accuracy is given for the following metrics: MAE = mean absolute error, RMSE = root mean squared error (unit: log CMC), MAPE = mean absolute percentage error (unit \%).}
	\label{tab:accuracies}
	\begin{tabular}{l | c  c c c| c  c  c c}
		& \multicolumn{4}{ c }{\textbf{Different temperature}} &  \multicolumn{4}{ c }{\textbf{Distinct surfactant}}\\
		& RMSE & MAE & MAPE & R$^2$ & RMSE & MAE & MAPE & R$^2$\\
		\hline
		GNN ensemble & 0.173 & 0.113 & 3.561 & 0.97  & 0.251 & 0.173 & 5.702 & 0.94 \\
	\end{tabular}
\end{table}

In the \emph{different temperature} scenario, the ensemble of GNNs exhibits an RMSE of 0.173, a MAE of 0.113, a MAPE of 3.561, and a high R$^2$ of 0.97. For the \emph{distinct surfactant} set, the ensemble of GNNs performs predictions on previously unseen surfactant molecules at single and multiple temperatures with an RMSE of 0.251, an MAE of 0.173, a MAPE of 5.702 and an R$^2$ of 0.94. Compared to the different temperature split, the errors slightly increase, which is expected since predicting the CMC of completely unseen surfactants is more difficult than predicting the CMC of a surfactant, for which CMC values at other temperatures have been included in training the models. 

\begin{figure}
\centering
\subfloat[Different temperature test split.]{\includegraphics[height = 7cm, width = 8cm]{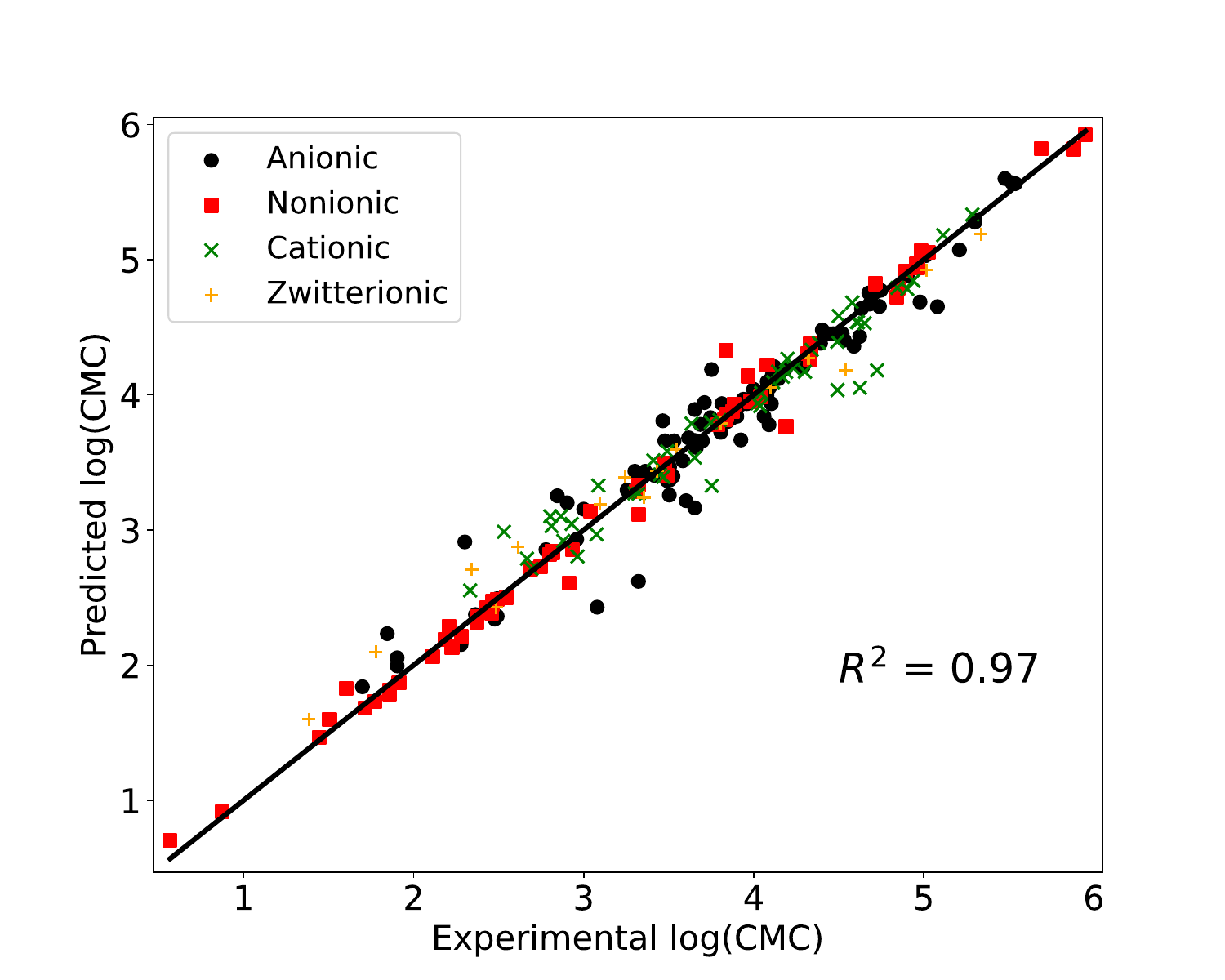}}
\hfill
\subfloat[Distinct surfactant test split.]{\includegraphics[height = 7cm, width = 8cm]{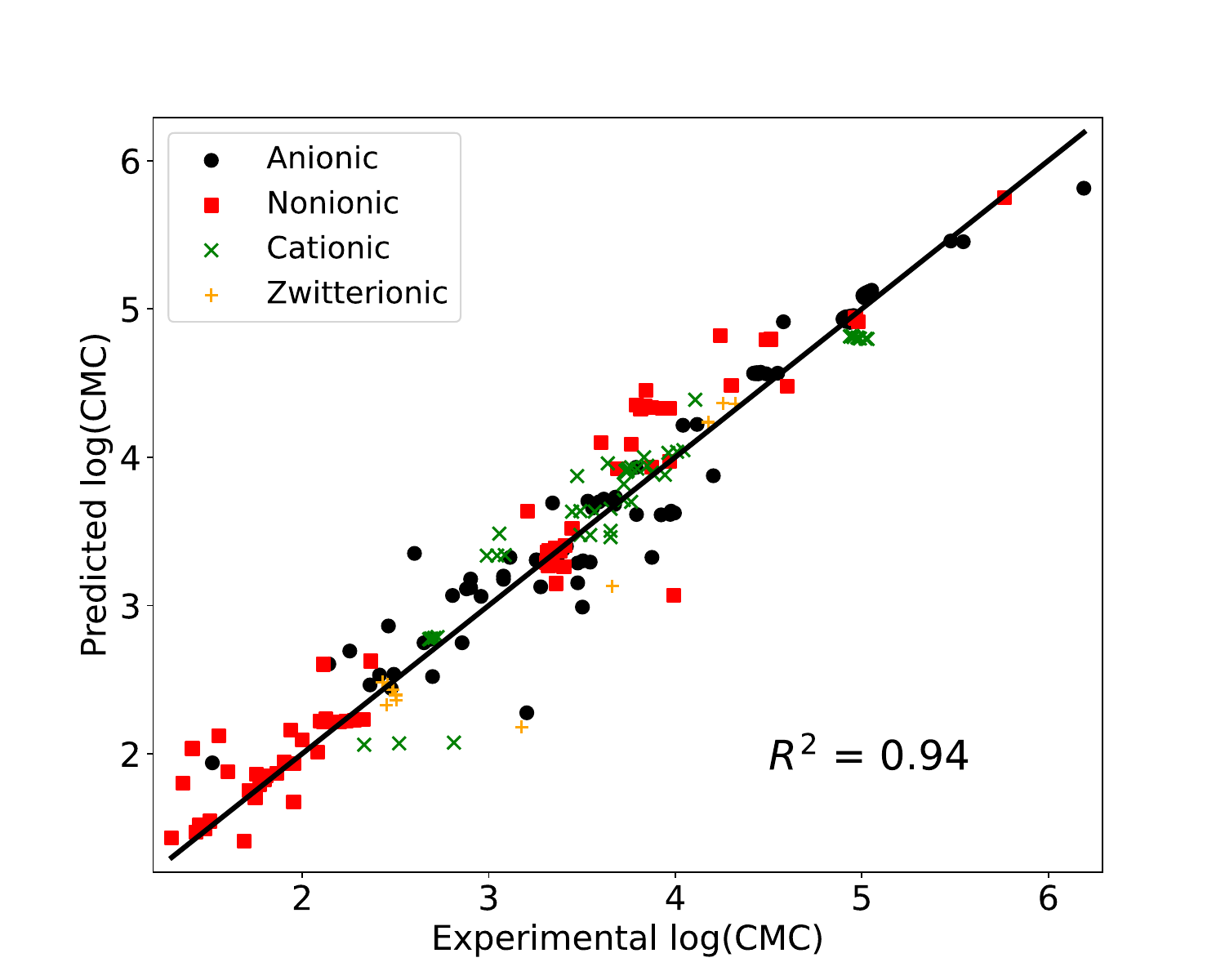}}
\caption{Parity plots of the ensemble of GNNs on the two test data sets: (a) different temperature and (b) distinct surfactant. Surfactant classes are highlighted with different colors and markers. The logarithm is applied to CMC in $\mu$M (base 10).}
\label{fig:parity_ensemble}
\end{figure}

Furthermore, we show parity plots for the predictions of the ensemble of GNN models for both test scenarios, different temperature and distinct surfactant, in Figure~\ref{fig:parity_ensemble}. For both scenarios, the models show very good agreement between measured and predicted data, as most of the points lie close to the diagonal, resulting in high R$^2$ scores of 0.97 and 0.94, respectively. We notice only a small number of outliers in both cases; the four points with the highest absolute error (AE) are reported in Figures~\ref{fig:apendix_type1_outliers} and~\ref{fig:appendix_type2_outliers}, respectively. The exact model predictions are directly compared with the experimental measurements in Tables~\ref{tab:appendix_outliers_distinct_temperature} and~\ref{tab:appendix_outliers_distinct_surfactant}.

Overall, the GNN ensembles provide a high predictive quality, also indicating generalization capabilities to new surfactants not included in the training.

\subsection{Comparison with previous works}\label{sec:prev_work}
\noindent
We further compare the predictive quality of our model to previously developed GNNs from the literature for CMC prediction. 
An overview of past model predictive performance is given in Table ~\ref{tab:work_comparison}.
Note that previous works have used different test sets, making a direct comparison difficult.
\cite{Qin2021} developed a GNN model for predicting CMCs of surfactant at constant temperature and reported an RMSE of 0.30 on their test set of 22 molecules.
This indicates a slightly worse performance on a less diverse test set compared to our GNN model with an RMSE of 0.25 evaluated on 218 different point with 50 distinct molecules.
In another recent work, \cite{Moriarty2023} analyzed the GNN model developed by \cite{Qin2021} and reported a combined GNN model with Gaussian Process (GP) with a test RMSE of 0.21, which is slightly lower than our findings. 
However, the test set was again limited to 22 molecules, which is only about 40\% of ours, and therefore may not represent the diversity of surfactant structures used in different practical applications. 
In their complementary data set, which contains 43 distinct surfactant molecules with some being outside the applicability domain of the model developed by \cite{Qin2021}, their best-performing model exhibited an RMSE of 1.32, which is almost six times higher than our findings.
In our recent work, we examined the performance of single- and multi-task learning on predicting the CMC and the surface excess concentration ($\Gamma_m$) of surfactant monomers with GNNs but without temperature dependency~\citep{brozos2024graph}. 
The multi-task GNN exhibited an RMSE of 0.31 and an R$^2$ of 0.94 for the CMC, which are 
slightly worse than our current results.
Therefore, we herein increase the predictive quality of GNNs for CMC prediction, although the presented prediction task includes the temperature dependency of the CMC and is thus presumably more challenging. 
We suspect, that the larger data set used in the present work (1,377 data points compared to 429 data points in our previous work~\citep{brozos2024graph}) positively influences the predictive capabilities of the GNNs. 
Overall, we find our GNN model to perform on a higher level (better test accuracy) with a broader applicability range, more diverse surfactant chemistry, than previous works. 

\begin{table}
    \centering
    \caption{Comparison of different GNN models for predicting the CMC of surfactants. Dash (--) indicates that values were not provided in previous studies.}
    \label{tab:work_comparison}
    \resizebox{\linewidth}{!}{%
    \begin{tabular}{c c c c c}
    \hline
    \textbf{Work} & \textbf{Test set size} & \textbf{Temperature dependency} & \textbf{RMSE} & \textbf{R$^2$}\\
    \hline
    \cite{Qin2021}& 22 & No & 0.30 & 0.91\\
    \cite{Moriarty2023} & 22 & No & 0.21  & -- \\ 
    \cite{Moriarty2023} - complementary set & 43 & No & 1.32 & -- \\ 
    Previous work~\citep{brozos2024graph} & 65 & No & 0.31 & 0.94 \\ 
    Current work & 218 & Yes & 0.25 & 0.94 \\
    \end{tabular}}
\end{table}

\subsection{Analysis of temperature impact on model accuracy}\label{sec:temp_accuracy}
\begin{figure}[ht]
    \centering
    \includegraphics[height = 8cm, width = 10 cm]{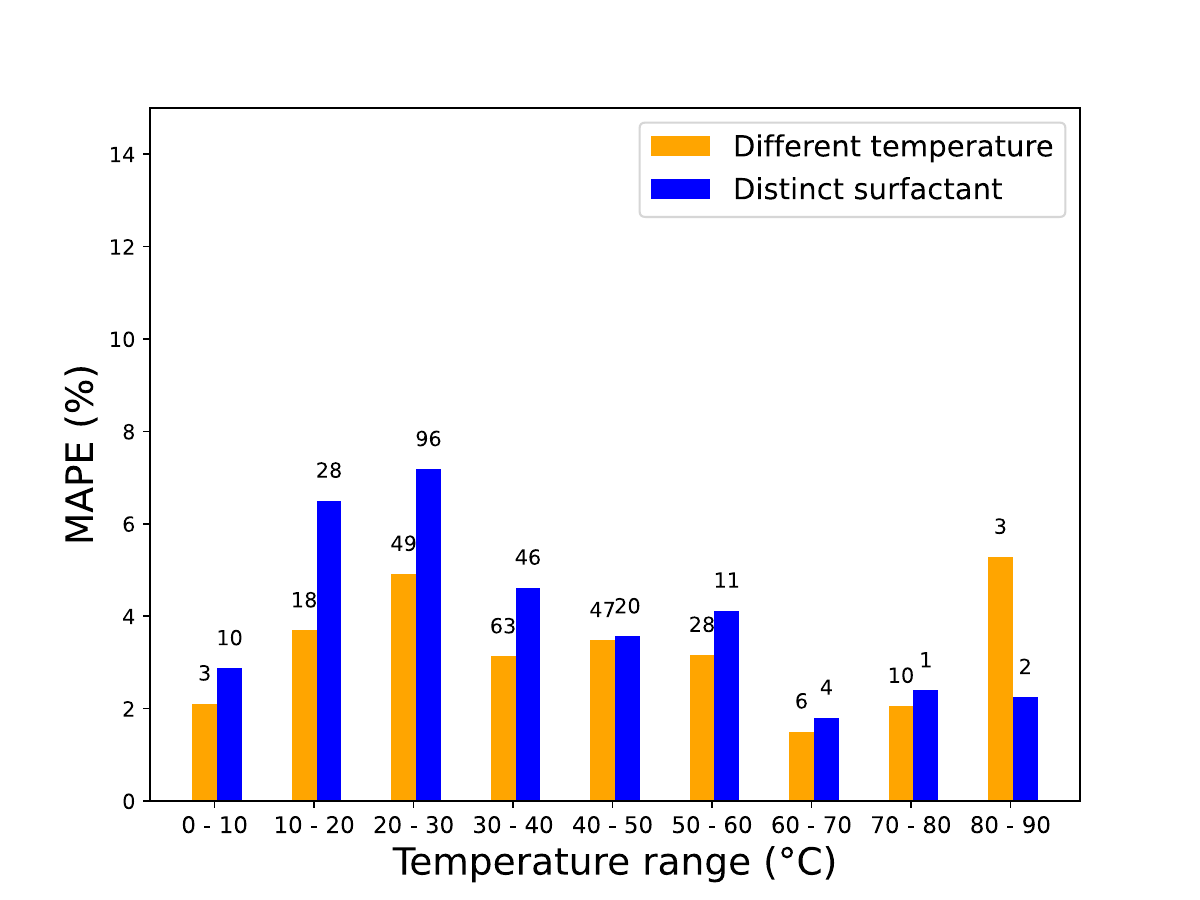}
    \caption{The mean absolute percentage error (MAPE) on the two test data sets: different temperature (yellow) and distinct surfactant (blue). The mentioned ranges exclude the left limit but include the right limit, for example (0 - 10] and (10 - 20]. The corresponding number of data points for each temperature range are denoted at the top of the respective bars.}
    \label{fig:mape_temperature}
\end{figure}

\noindent
We further examine the model performance for both test sets, different temperature and distinct surfactant, in the whole temperature spectrum. Therefore, we separate the individual temperatures into temperature bins (of 10$^\circ$C ranges) and we calculate the MAPE in each one.
The results are plotted in Figure~\ref{fig:mape_temperature}. 
In most temperature bins, the MAPE is higher on the distinct surfactant split than on the new temperature split, which is expected due to the more challenging prediction task (cf. Section~\ref{sec:pred_new_temperatures}). We do not observe any temperature range with a significant high error in both test scenarios. As expected, the highest amount of measurements are between 20$^\circ$C and 30$^\circ$C.

\begin{figure}[hbp]
\centering
\begin{subfigure}[c]{0.49\textwidth}
	\centering
	\includegraphics[trim={0cm 5.5cm 0cm 8cm},clip,width=\textwidth]{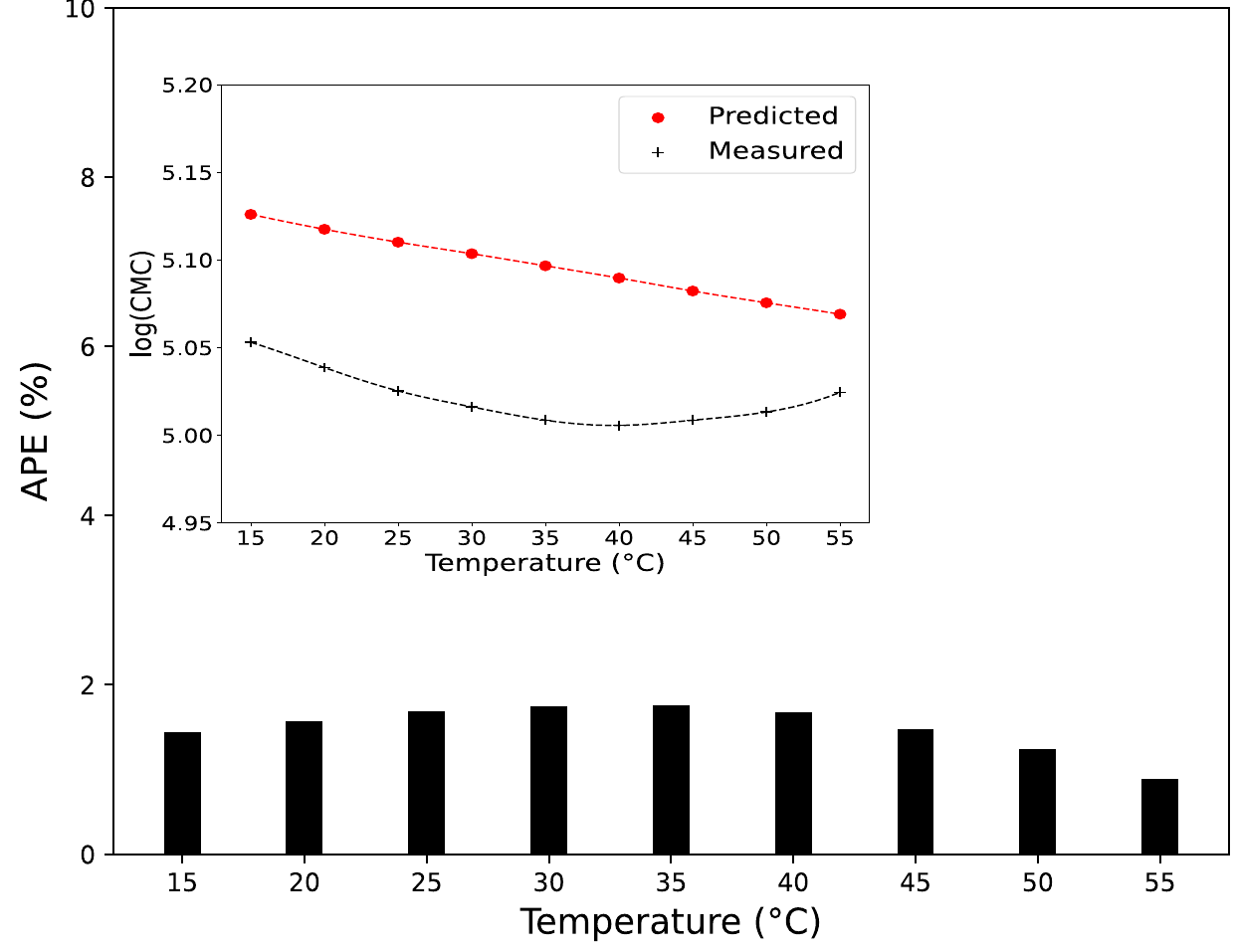}
	\subcaption{Sodium decanoate~\citep{Blanco2005}.}
\end{subfigure}
\begin{subfigure}[c]{0.49\textwidth}
	\centering
	\includegraphics[trim={0cm 6cm 0cm 7.5cm},clip,width=\textwidth]{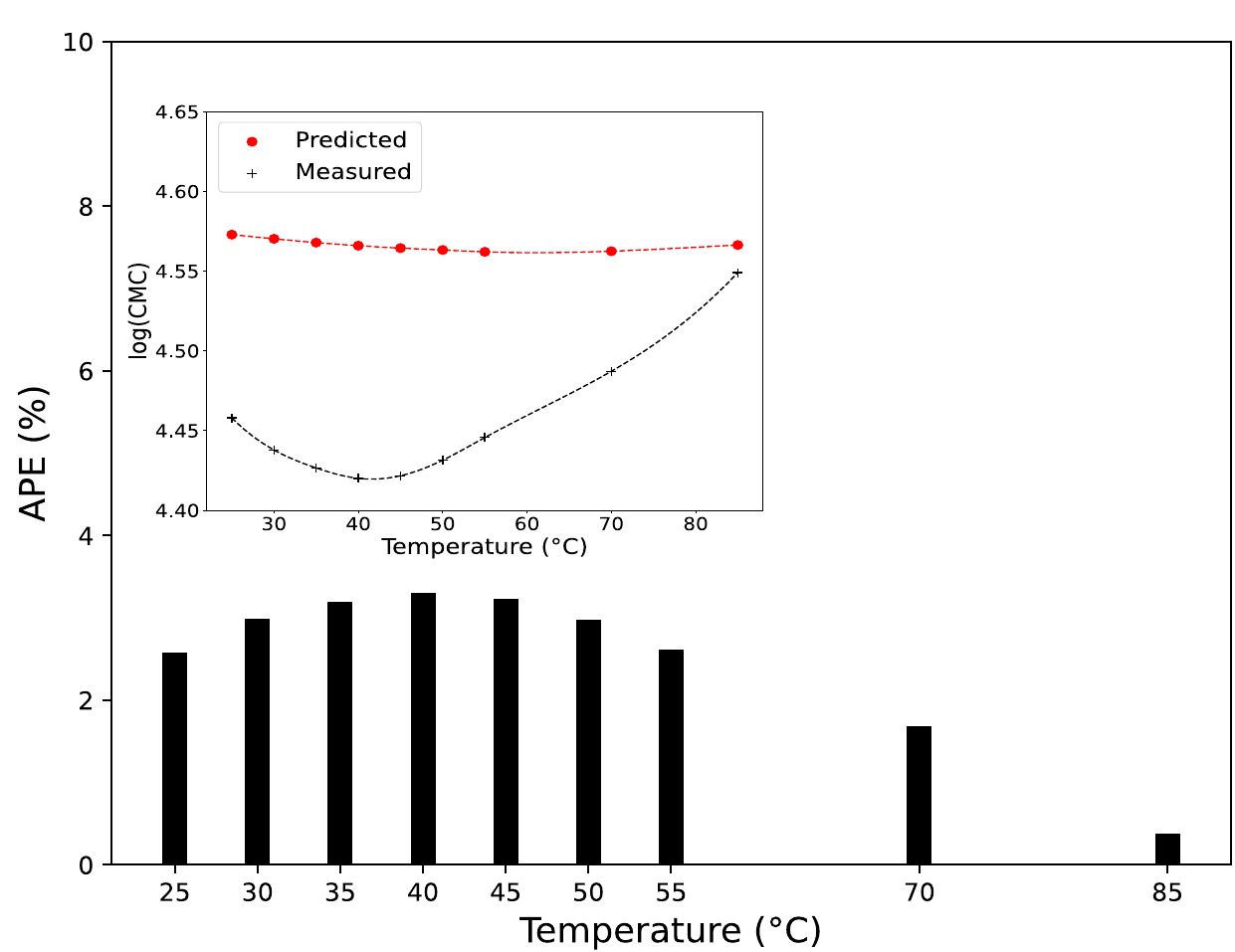}	
	\subcaption{Potassium perfluorooctanoate~\citep{Muzzalupo1995,Mukerjee1971}.}
\end{subfigure}
\begin{subfigure}[c]{0.49\textwidth}
	\centering
	\includegraphics[trim={0cm 6cm 0cm 7.5cm},clip,width=\textwidth]{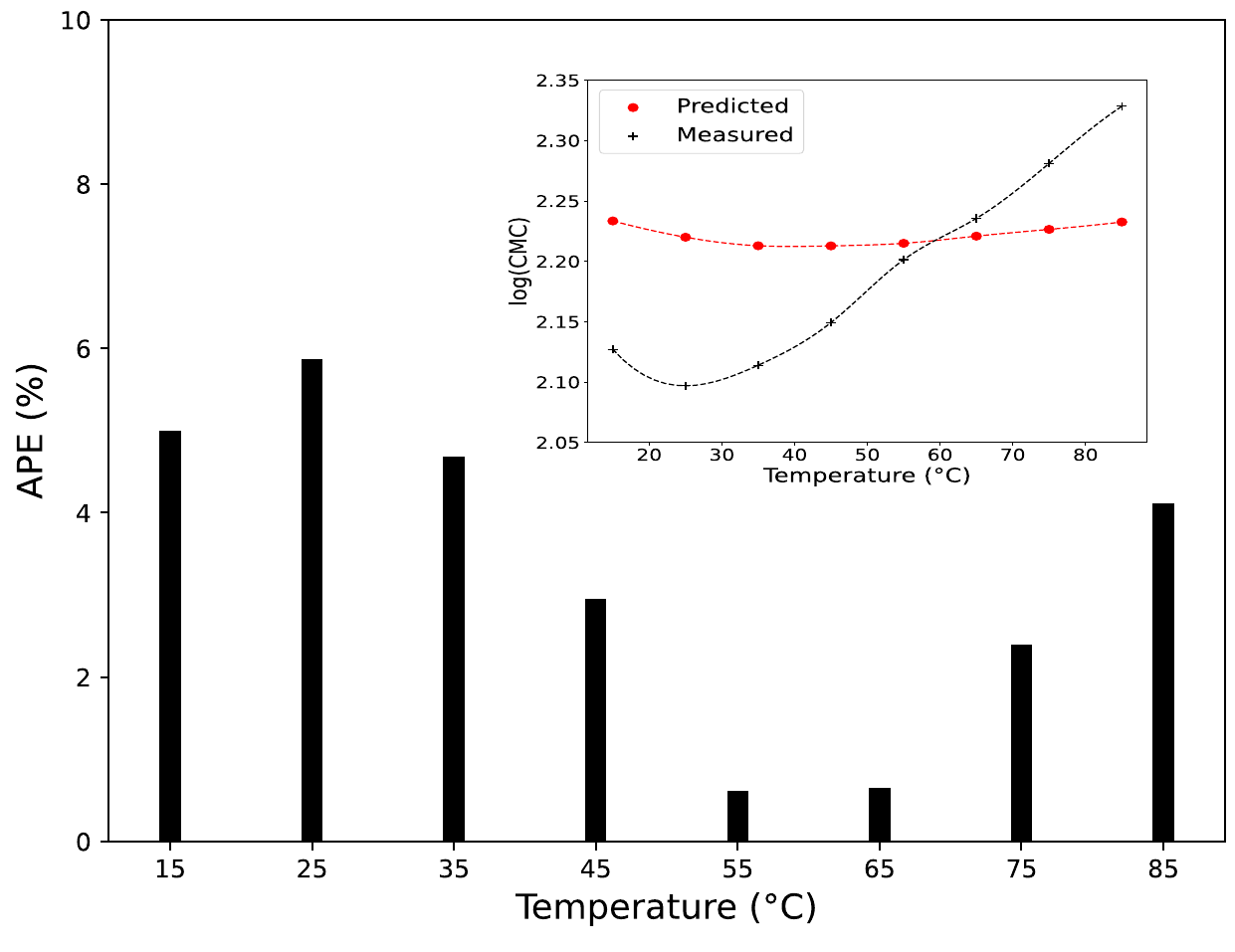}
	\subcaption{Octoxynol 4~\citep{Mukerjee1971}.}
\end{subfigure}
\begin{subfigure}[c]{0.49\textwidth}
	\centering
	\includegraphics[trim={0cm 6cm 0cm 7.5cm},clip,width=\textwidth]{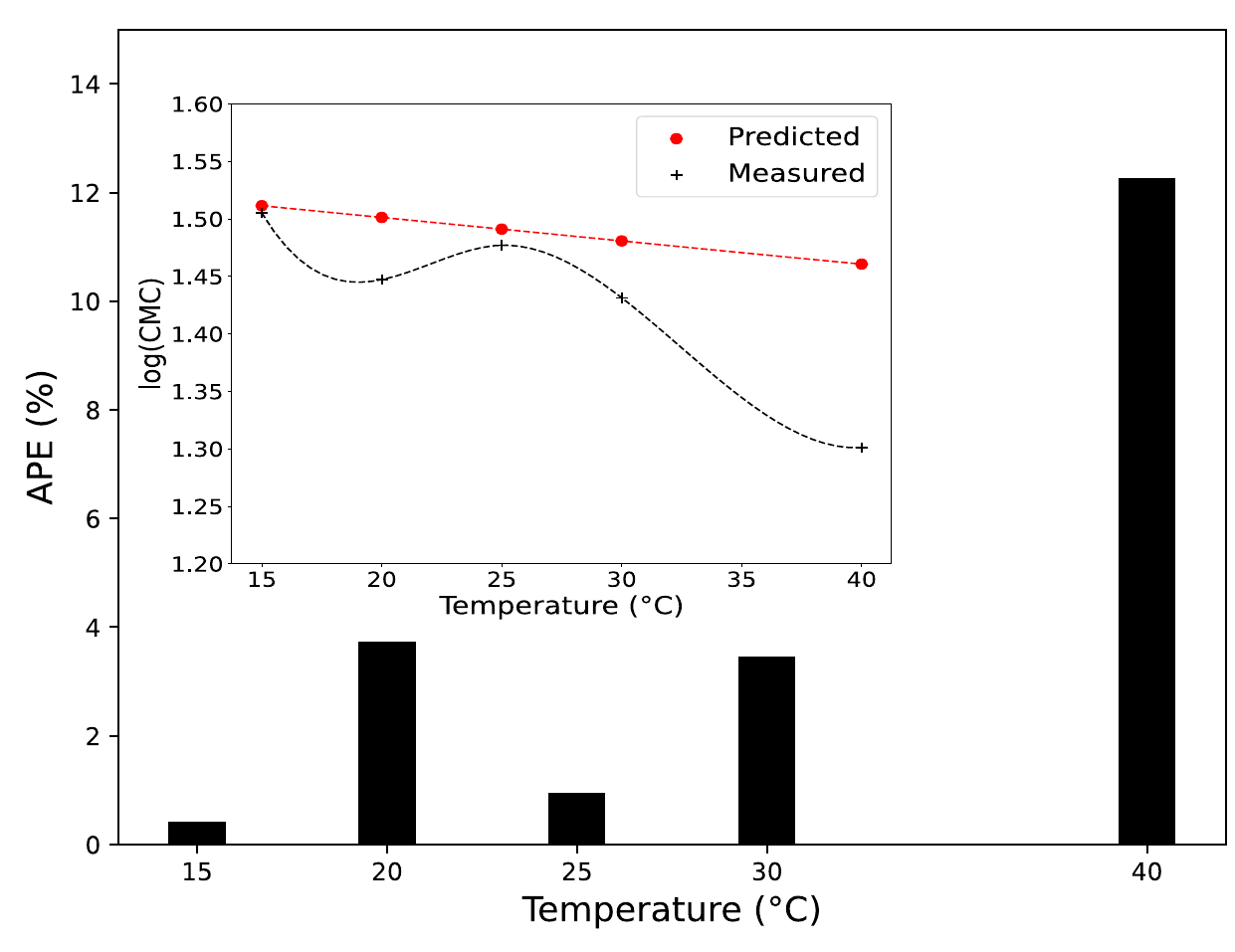}
	\subcaption{C$_{13}$E$_{8}$~\citep{Meguro1981,Schick1987}.}
\end{subfigure}
\caption{The GNN predicted versus experimental temperature-dependent performance on four surfactants, part of our test set is shown. Predicted and measured log CMC values are plotted together and connected through a cubic interpolation in the corresponding insets. The logarithm is applied to CMC in $\mu$M.}
\label{fig:surfs_temperature_examples}
\end{figure}

We further visualize the temperature dependence of the GNN predicted CMC versus the experimental CMC and the Absolute Percentage Error (APE) for four randomly selected surfactants from the distinct surfactant test set in Figure~\ref{fig:surfs_temperature_examples}. 
Please note that different scales of CMC are used in the plots since the order of CMC magnitude varies for different surfactants (cf. Section~\ref{sec:data_sets} and Appendix~\ref{sec:appendix_A}). The scale of APE in Figure~\ref{fig:surfs_temperature_examples}d also differs from the rest.
Given this wide range of CMC values, Figure~\ref{fig:surfs_temperature_examples} shows that the model accurately predicts the order of magnitude of CMC at multiple temperatures.
The model also identifies trends for the temperature dependency of the CMC for most cases, e.g., the U-shaped relationship for the two surfactants in Figures~\ref{fig:surfs_temperature_examples}b-c.
For sodium decanoate shown in Figure~\ref{fig:surfs_temperature_examples}a, the decrease in the CMC for the range of 15-40$^\circ$C is predicted correctly, but the increase of the CMC up to 55$^\circ$C is not captured. The highest APE (12\%) is observed at Figure~\ref{fig:surfs_temperature_examples}d at 40$^\circ$C.
While general trends are mostly captured, the model seems to underestimate the temperature effect on the CMC. The lack of high sensitivity to temperature changes may due to the fact that for some surfactants only marginal changes in CMC value are observed at different temperatures, as shown in Figure~\ref{fig:descriptive_examples}, which are even less pronounced on the logarithmic scale used for model training.
Thus, additional training data and further model refinement would be desirable to fully capture the detailed temperature effects in future work.

\subsection{Predictive performance per surfactant class}\label{sec:performance_class}
\noindent
As discussed previously (cf. Section~\ref{sec:intro}), the temperature effect of the CMC is unique for surfactants belonging to different surfactant classes. Thus, from a product development viewpoint, analyzing the results for each class is highly interesting. For example, one would want to identify surfactants and classes that exhibit a lowering of CMC with temperature, since the minimum CMC value and the corresponding temperature may be valuable for different applications (e.g., in laundry applications).

\begin{figure}
    \centering
    \includegraphics[height = 8cm, width = 10 cm]{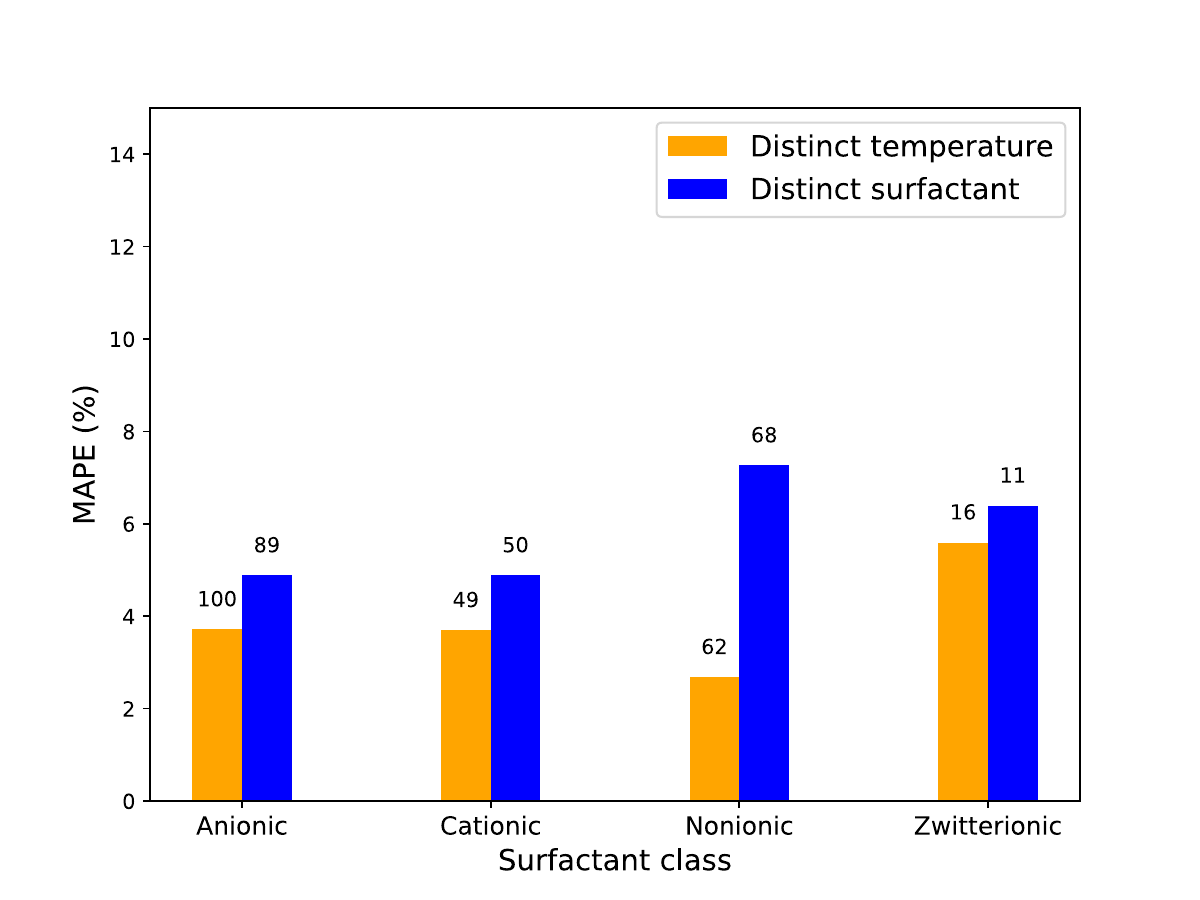}
    \caption{The mean absolute percentage error (MAPE) per surfactant class on the two test data sets: different temperature (yellow) and distinct surfactant (blue). The corresponding number of data points for each temperature range are denoted at the top of the respective bars.}
    \label{fig:mape_class}
\end{figure}

We calculate the MAPE per surfactant class for both test scenarios and report the results in Figure~\ref{fig:mape_class}. Overall, all surfactant classes demonstrate a MAPE $< 8\%$. Ionic surfactants have a slightly higher error on the distinct surfactant split in comparison to the different temperature split, as unseen surfactant structures are introduced on the test set (cf. Section~\ref{sec:split_types}). Specifically, anionic and cationic surfactants have the lowest errors in total for the distinct surfactant split. Since they generally exhibit a U-shaped relationship, the model is able to make accurate predictions. Zwitterionics, do not always exhibit the same general CMC-temperature relationship, and therefore decreased model accuracy is observed. For nonionic surfactants, the MAPE is more than doubled in the distinct surfactant split. Here, sugar-based surfactants are also considered. The high complexity of CMC dependency on temperature for such surfactants decreases model accuracy. The model performance on sugar-based surfactants is investigated in the next subsection of this work. Overall, we find model performance to be highly correlated with the number of possible CMC-temperature relationships (U-shaped, monotonically decrease, etc.) in each surfactant class. 

\subsection{Predictive performance on sugar-based surfactants}\label{sec:sugar_results}
\noindent

Bio-based surfactants and biosurfactants that are comprised of sustainable and natural-based feedstock are of technological importance to personal and home care industries. These often contain either single or multiple sugar groups, which control surfactant properties like foaming, cleansing, etc. Such sugar-based surfactants are complex molecules with many chiral centers on the sugar head. Therefore different anomeric configurations of the sugar groups and hence the entire surfactant structure are possible. In GNNs operating on the molecular graph, which is a 2D representation of the molecule (cf. Section~\ref{sec:graph_neural}), such spatial/geometric information is typically solely captured as an atomic feature, thus making it challenging to accurately predict properties of complex sugar-based surfactants. Therefore, we analyze the performance of the model on some selected sugar-based surfactants present in the \emph{distinct surfactant} test set. 

\begin{figure}
	\begin{subfigure}[c]{0.49\textwidth}
		\centering
		\includegraphics[trim={0cm 7cm 0cm 7cm},clip,width=\textwidth]{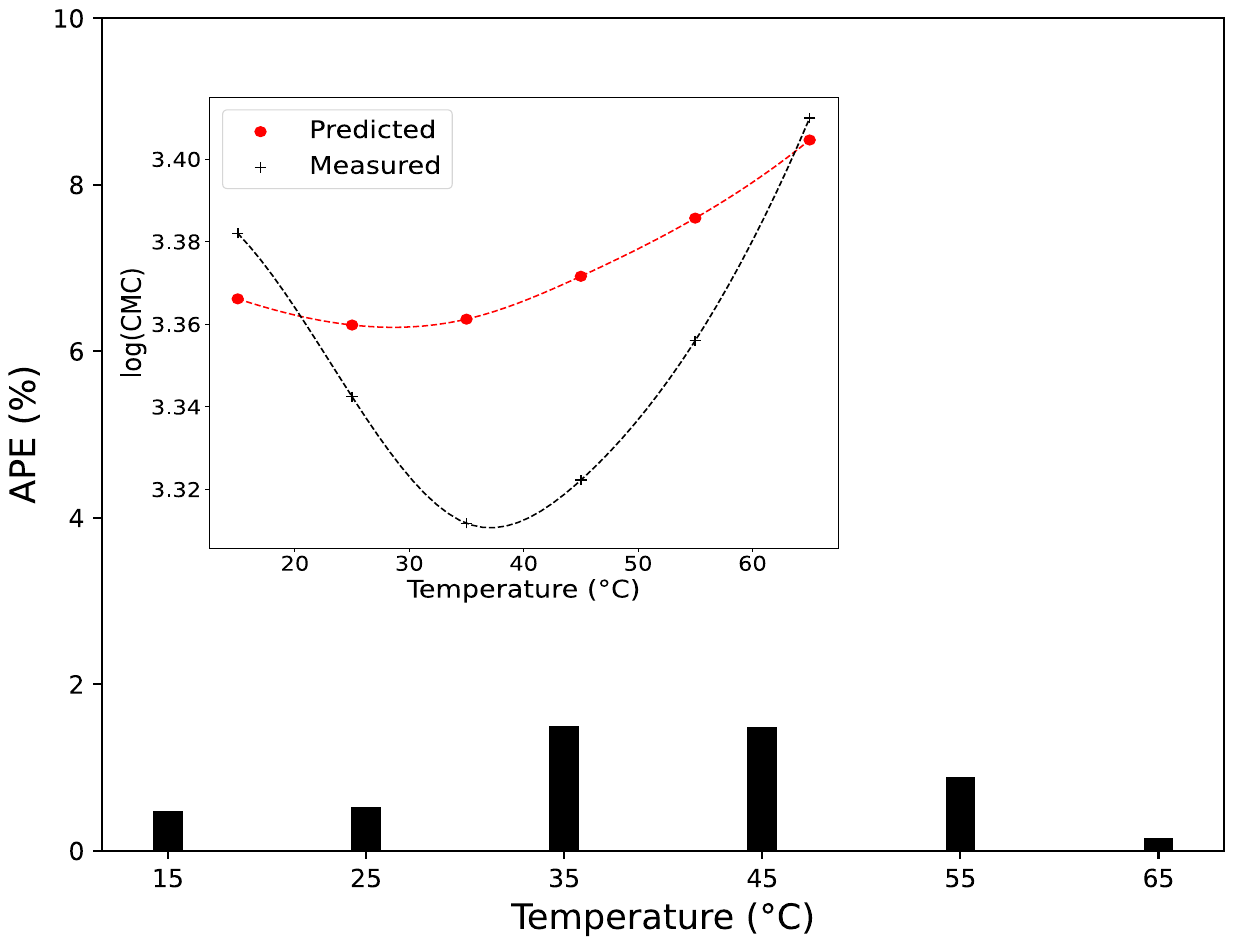}
		\subcaption{Decyl $\beta$-D-maltoside~\citep{Angarten2014}.}
	\end{subfigure}
	\begin{subfigure}[c]{0.49\textwidth}
		\centering
		\includegraphics[trim={0cm 6cm 0cm 8cm},clip,width=\textwidth]{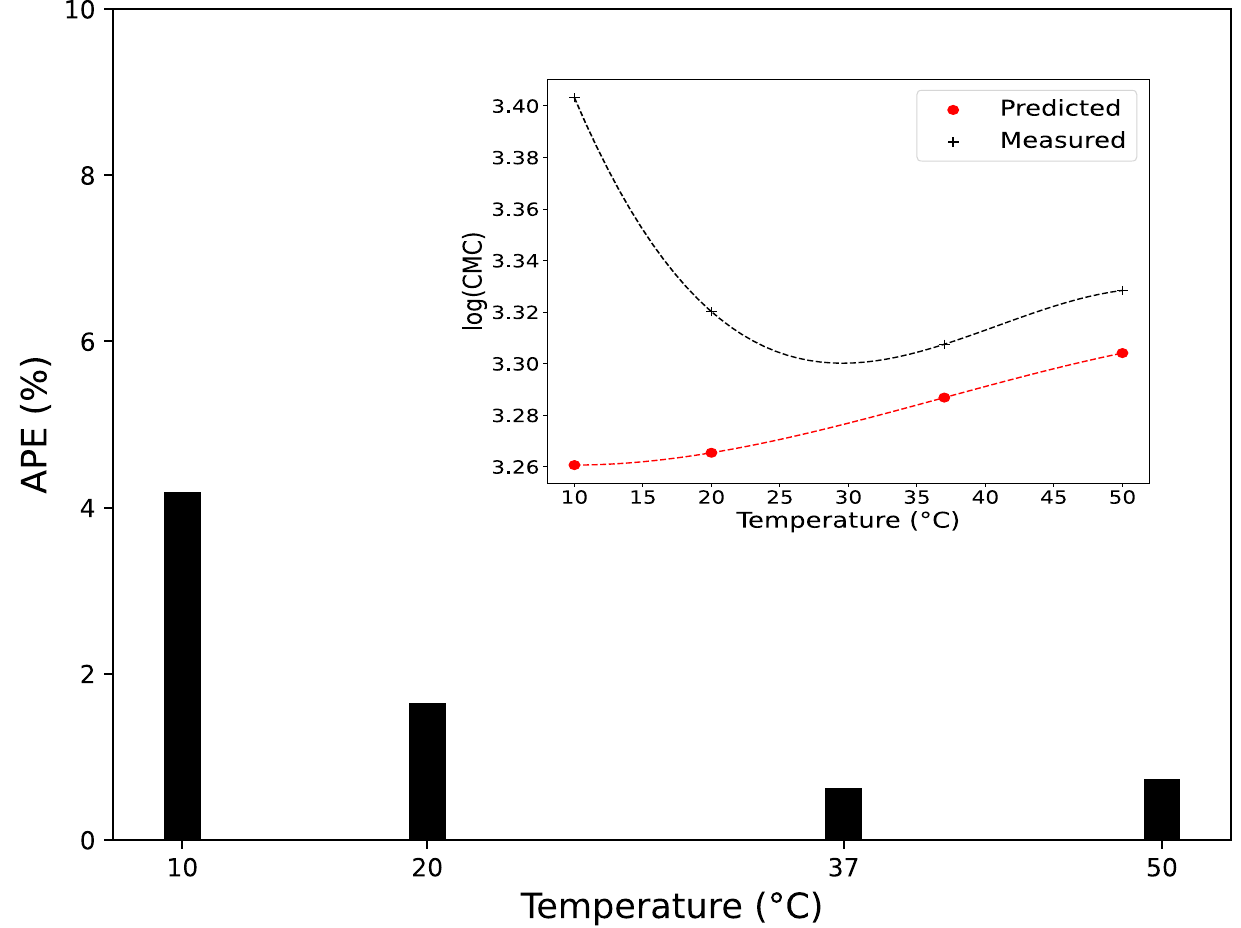}
		\subcaption{Decyl $\beta$-D-glucoside~\citep{Castro2018}.}
	\end{subfigure}
	\begin{subfigure}[c]{\textwidth}
		\centering
		\includegraphics[trim={0cm 5.5cm 0cm 7.5cm},clip,width=0.49\textwidth]{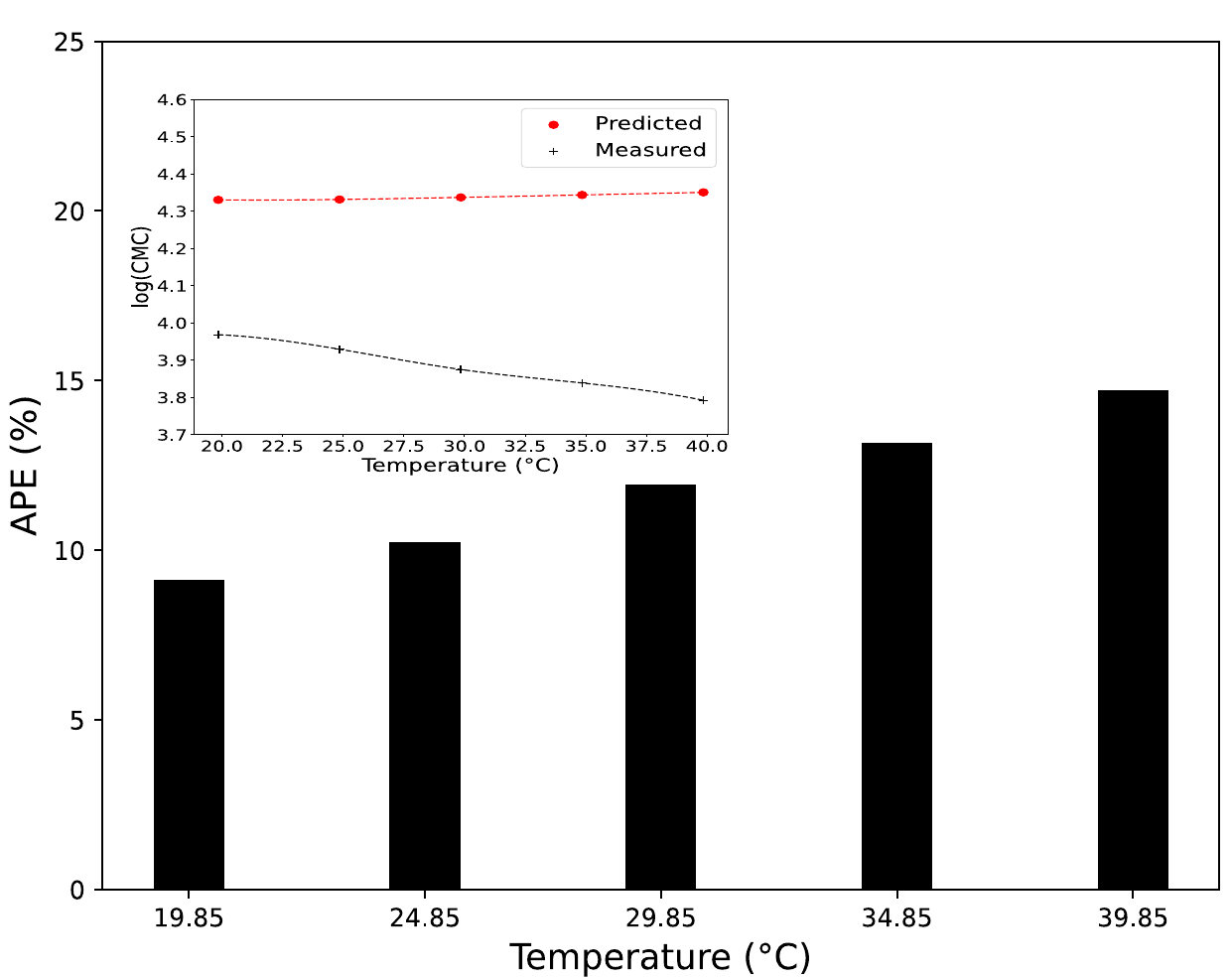}
		\subcaption{Octyl $\beta$-D-Thioglucoside~\citep{MolinaBolivar2004}.}
	\end{subfigure}
	\caption{Ensembled predictions on sugar-based surfactant in the distinct surfactant test set. Predicted and measured log CMC are plotted together and connected through cubic interpolation. The logarithm is applied to CMC in $\mu$M. The absolute percentage error (APE) per temperature is also illustrated.}
	\label{fig:sugar_based}
\end{figure}

We plot the temperature-dependent CMC predictions together with the experimental CMC values from the literature, as well as the APE per temperature in Figure~\ref{fig:sugar_based}. Since absolute (not logarithmic) CMC values are also of interest in practical, industrial applications, we report the absolute predictions on the three sugar-based surfactants in Table~\ref{tab:appendix_sugar_based_predictions}.

We observe the ability of our model to accurately predict the magnitude of the CMC values for two out of three sugar-based surfactants, namely decyl $\beta$-D-glycoside and decyl $\beta$-D-maltoside. This is evident by the low APE for these two surfactants across all temperatures. Furthermore, the correct trend of the temperature dependency is identified, except for decyl $\beta$-D-maltoside at 10$^\circ$C. The reason may be the low amount of measurements at this temperature, as shown above in Figure~\ref{fig:temperature_distribution}. On the other hand, our model fails to accurately predict the real CMC values of octyl $\beta$-D-Thioglucoside. As noted by \cite{Gaudin2019}, the nature of the linker, i.e., the connection between the carbohydrate and the alkyl chain, impacts the surfactant properties and consequently the CMC. Replacing the ether -O- with thioether -S- linker, leads to incorrect predictions from our model. We note that our data set contains no other sugar-based surfactant with thioether as the linker. Thus, expanding the data set further with more experimental data on sugar-based surfactants would improve the predictions of our model.

We are further interested in the effect of anomeric configuration and chirality centers of sugar-based surfactants on the CMC predictions. Since all previously discussed sugar-based surfactants had an $\beta$ anomeric configuration, we analyze the model performance on heptyl-$\alpha$-D-mannoside and heptyl-$\alpha$-D-galactoside at a single temperature. The calculated and measured CMC values are given in Table~\ref{tab:sugar_results}. We observe a relatively high absolute error for both surfactants, namely 0.306 and 0.283 respectively. Note here that the overall MAE was found to be 0.173 (cf. Section~\ref{sec:pred_new_temperatures}). 
The most often present sugar-based surfactants on our database are of the n-alkanoyl-$\beta$-D-glucoside type. In comparison with them, both surfactants exhibit structural differences in their anomeric configuration and chirality centers of the sugar head. Specifically, our training data set contains only two surfactants with mannoside and three surfactants with galactoside as the polar head. Given the limited amount of similarly structured surfactants in the training set, the model provides acceptable CMC predictions but with an above-average error compared to the overall MAPE of the distinct surfactant test set. 
To enable the GNN to additionally learn different structures/linkers in sugar-based surfactants, systematic CMC data on these surfactants should be provided in the future.

\begin{table}
\centering
\caption{Comparison between predicted and experimental CMC values for two sugar-based surfactants. The CMC values outside the parenthesis given are the logarithmic ones. The CMC values inside the parenthesis represent the absolute value and are given in mM. Both surfactants are measured at 25$^\circ$C.}
\label{tab:sugar_results}
\resizebox{\linewidth}{!}{%
\begin{tabular}{l | c |  c | c }
Surfactant &  \textbf{Predicted CMC ($\mu$M)}  & \textbf{Measured CMC ($\mu$M)} & \textbf{Source}\\
\hline
Heptyl $\alpha$-D-mannoside & 4.792 (61.937)  & 4.49 (30.6) & ~\citep{Gaudin2019}\\
Heptyl $\alpha$-D-galactoside & 4.795 (62.373) & 4.512 (32.5) & ~\citep{Gaudin2019}\\
\end{tabular}}
\end{table}

\section{Conclusion}\label{sec:conclusion}
\noindent 

We develop a GNN model for the prediction of the temperature-dependent CMC of surfactants. 
For this, we first assemble a data set with 1,377 CMC measurements at multiple temperatures ranging from 0$^\circ$C to 90$^\circ$C for 492 unique surfactants covering all surfactant classes, i.e., anionic, cationic, zwitterionic, and nonionic surfactants.
We then use this data set to train an ensemble of GNN models directly on the molecular graph of the surfactants and the temperature information, thereby enabling the prediction of temperature dependency of the CMC, a key interest in personal and home care applications.

The GNN model exhibits very high predictive quality.
Specifically, the model shows accurate predictions for surfactants included in the training but at a different temperature and for surfactants not seen during training at all, thus generalizing to new surfactants.
Model performance remained constant throughout all temperature ranges but varied per surfactant class.
The GNN model also outperforms previously developed data-driven models for predicting the CMC at constant temperature. 
Detailed temperature dependencies are not fully captured by the model yet, i.e., the model underestimates the sensitivity of the CMC to changes in the temperature, demonstrating the need for additional temperature-dependent CMC data and model refinement. 

\par We specifically investigate sustainably sourced sugar-based surfactants. 
The GNN model incorporates anomeric and chirality information into the feature vector and can thus distinguish between different anomers and isomers. 
The GNN provides accurate predictions of the CMC at multiple temperatures for two exemplary sugar-based surfactants but also shows limitations in differentiating ether (-O-) and thioether (-S-) linker in a third selected sugar-based surfactant. Further evaluation of the anomeric and chirality information effect on the model performance showed an above-average error and model limitations in distinguishing between different chiral centers on the sugar head. 
Yet, the model accurately predicts the order of magnitude of the temperature-dependent CMC values for sugar-based surfactants.

\par Future work should focus on acquiring additional temperature-dependent CMC data for model training. To account for the CMC relation to the geometry of the molecular structure, which is particularly relevant for sugar-based surfactants, geometric GNNs that consider atomic coordinates could be investigated. Also, extending the GNN model to incorporate pH conditions and defining the validity domain of the model would be insightful extensions. Finally, obtaining chemical intuition on how the model performs predictions by using explainable AI methods could help to better understand surfactant self-aggregation.

\section*{Acknowledgments}

\noindent The BASF authors (C. Brozos, S. Bhattacharya, E. Akanny, and C. Kohlmann) were funded by the BASF Personal Care and Nutrition GmbH.
J. G. Rittig and A. Mitsos acknowledge funding from the Deutsche Forschungsgemeinschaft (DFG, German Research Foundation) – 466417970 – within the Priority Programme ``SPP 2331: Machine Learning in Chemical Engineering''.
Additionally, J. G. Rittig acknowledges the support of the Helmholtz School for Data Science in Life, Earth and Energy (HDS-LEE).

\section*{Data availability}
\noindent All Python scripts and the test data used in this work are available as open-source in our~\href{https://github.com/brozosc/Predicting-the-Temperature-Dependence-of-Surfactant-CMCs-using-Graph-Neural-Networks}{GitHub repository}.

\section*{Author contribution}
\noindent
\textbf{Christoforos Brozos}: Conceptualization, Methodology, Software, Data curation, Validation, Formal analysis, Writing - Original Draft, Writing - Review \& Editing, Visualization

\noindent\textbf{Jan G. Rittig}: Conceptualization, Methodology, Software, Formal analysis, Writing - Review \& Editing

\noindent\textbf{Sandip Bhattacharya}: Conceptualization, Methodology, Formal analysis, Supervision, Writing - Review \& Editing

\noindent\textbf{Elie Akanny}: Conceptualization, Methodology, Experimental methodology \& measurements, Writing - Review \& Editing

\noindent\textbf{Christina Kohlmann}: Writing - Review \& Editing, Supervision, Funding acquisition

\noindent\textbf{Alexander Mitsos}: Writing - Review \& Editing, Supervision, Funding acquisition

  \clearpage

  \bibliographystyle{apalike}
  \renewcommand{\refname}{Bibliography}
  \bibliography{literature.bib}

\cleardoublepage

\renewcommand{\thefigure}{A\arabic{figure}}

\appendix
\section{Figures}\label{sec:appendix_A}
\noindent 
\begin{figure}[h!]
    \centering
    \includegraphics[height = 8 cm, width = 12 cm]{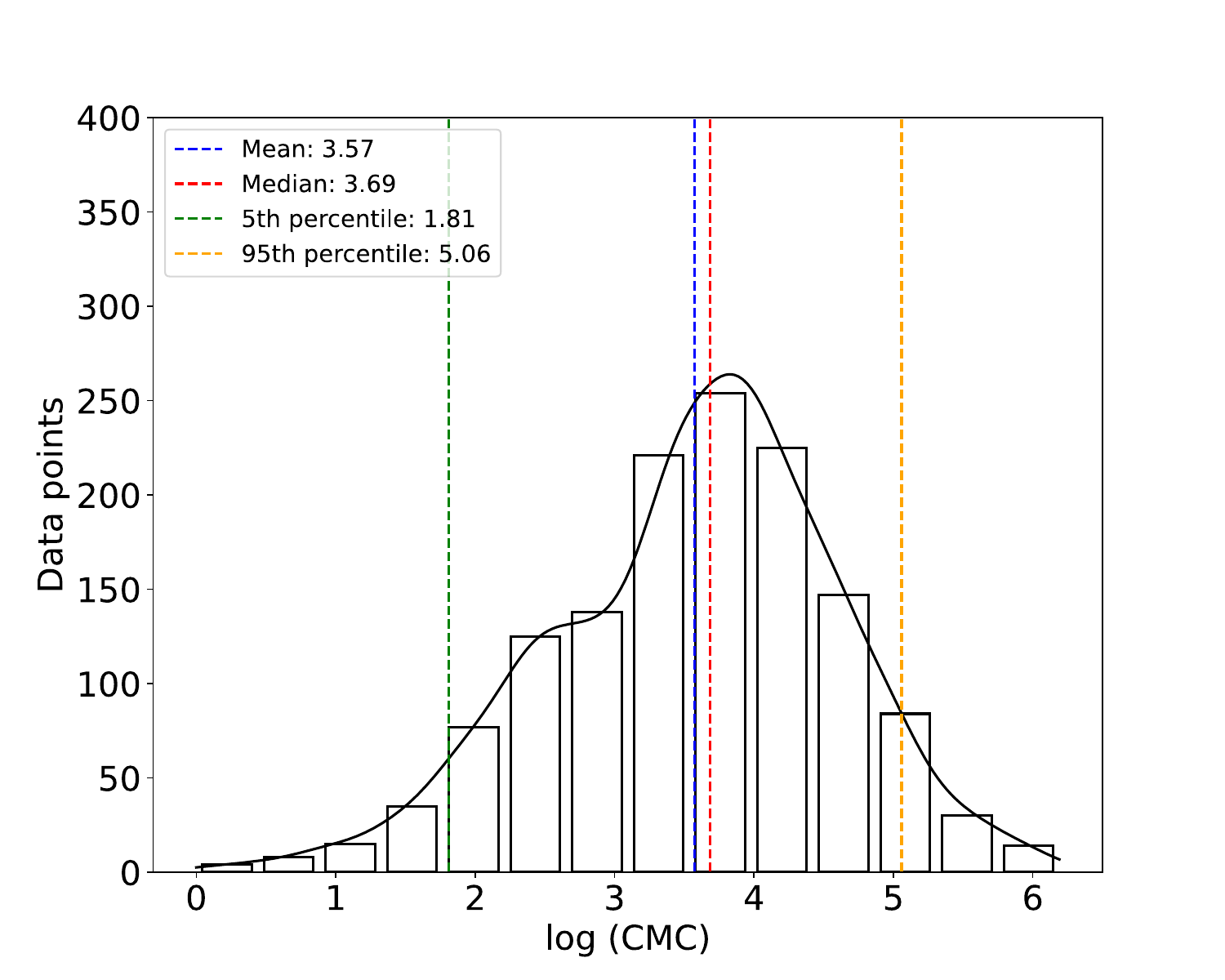}
    \caption{Statistical overview of the CMC. The logarithm is applied to CMC in $\mu$M.}
    \label{fig:apendix_cmc_distribution}
\end{figure}

\begin{figure}[htpb]
\centering
\subfloat[]{\includegraphics[height = 1.1cm, width = 7cm]{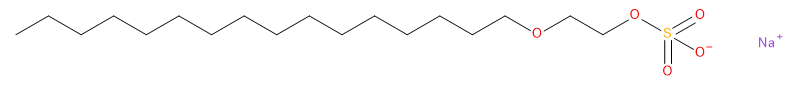}}
\hfill
\subfloat[]{\includegraphics[height = 1.1cm, width = 7cm]{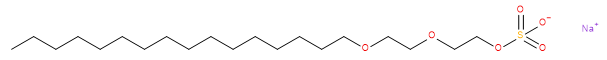}}
\hfill
\subfloat[]{\includegraphics[height = 1.1cm, width = 7cm]{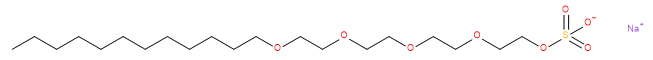}}
\hfill
\subfloat[]{\includegraphics[height = 1.1cm, width = 6cm]{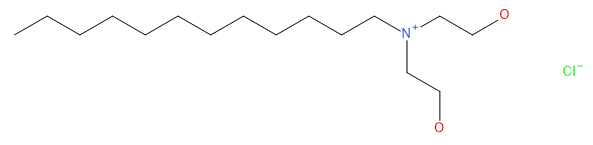}}
\caption{Outliers of the ensembled GNN model in different temperature split. The first three of them belong to anionic class and the last one is cationic.}
\label{fig:apendix_type1_outliers}
\end{figure}

\begin{figure}[htpb]
\centering
\subfloat[]{\includegraphics[height = 1.1cm, width = 6cm]{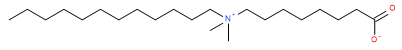}}
\hfill
\subfloat[]{\includegraphics[height = 1.1cm, width = 6cm]{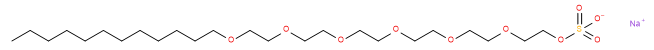}}
\hfill
\subfloat[]{\includegraphics[height = 1.2cm, width = 6cm]{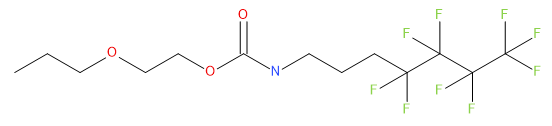}}
\hfill
\subfloat[]{\includegraphics[height = 1.3cm, width = 6cm]{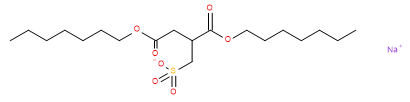}}
\caption{Outliers of the ensembled GNN model in different distinct surfactant split. The first outlier belongs to zwitterionic class, the third in nonionic class and the other two to the aninionic class.}
\label{fig:appendix_type2_outliers}
\end{figure}

\cleardoublepage
\renewcommand{\thetable}{B\arabic{table}}

\section{Tables}
\noindent 
\begin{table}[h!]
	\begin{center}
		\caption{Atom features used in the molecular graph representation. All features are implemented as one-hot-encoding.}
		\label{tab:appendix_node_features}
		\begin{tabular}{c | p{7cm} | c}
		\textbf{Feature} & \centering \textbf{Description} & \textbf{Dimension}\\
		\hline
		atom type & \centering atom type (C, N, O, S, F, Cl, Br, Na, I, B, K, H, Li) & 13\\
        is in a ring & \centering if the atom is part of a ring & 1 \\
		is aromatic &\centering if the atom is part of an aromatic system & 1 \\
		hybridization &\centering sp$^2$, sp$^3$ & 2 \\
        chirality &\centering unspecified, clockwise, counter clockwise & 3 \\
        charge &\centering formal charge of the atom (-1,0,1) & 3 \\
		\# bonds &\centering number of bonds the atom is involved in & 5 \\
		\# Hs &\centering number of bonded hydrogen atoms & 5 \\
		\hline
		\textbf{Total} & & \textbf{33}\\
		\end{tabular}
	\end{center}
\end{table}

\begin{table}[htpb]
	\begin{center}
		\caption{Edge features used in the molecular graph representation. All features are implemented as one-hot-encoding.}
		\label{tab:appendix_edge_features}
		\begin{tabular}{c | c | c}
		\textbf{Feature} & \textbf{Description} & \textbf{Dimension}\\
		\hline
		bond type & single, double, or atomatic  & 3 \\
		is in a ring & is the bond part of a ring ? & 1\\
		conjugated & is the bond conjugated ? & 1 \\
		stereo & none or E/Z & 3 \\
		\hline
		\textbf{Total} & & \textbf{8}\\
		\end{tabular}
	\end{center}
\end{table}

\begin{table}
    \begin{center}
    	\caption{Hyperparameters of the GNN model investigated through a grid search. The hyperparameter dimensions refers to the size of the molecular fingerprint and the size of the MLP.}
    	\label{tab:appendix_model_hyperparameters}
        \begin{tabular}{c|c |c}
             \textbf{Hyperparameter}& \textbf{Range} & \textbf{Optimized geometry}\\
             \hline
             Graph convolutional layers & (1, 2) & \textbf{1}\\
             Graph convolutional type & (NNConv, GINEConv) & \textbf{GINEConv}\\
             Usage of GRU & (True, False)& \textbf{False} \\
             Initial learning rate & (0.005, 0.01, 0.05) & \textbf{0.005} \\
             Batch size & (16, 32, 64) & \textbf{32}\\
             Dimensions & (64, 128) & \textbf{128}\\
             Number of MLP layers & 3 & \textbf{3} \\
             Activation function & ReLU & \textbf{ReLU} \\
             Maximum epochs & 300 & \textbf{300} \\
             Early stopping patience & 60 & \textbf{60} \\
             Learning rate decay & 0.8 & \textbf{0.8} \\
             Patiance & 3 & \textbf{3} \\
             Optimizer & Adam & \textbf{Adam}\\
        \end{tabular}
    \end{center}
\end{table}

\begin{table}
	\begin{center}
		\caption{Comparison of predicted vs measured CMC values of the 4 outlier for the different temperature split. The data source is provided.}
		\label{tab:appendix_outliers_distinct_temperature}
		\resizebox{\linewidth}{!}{%
		\begin{tabular}{l | c c c c}
		& \multicolumn{4}{ c }{\textbf{Different temperature test set}} \\
        &  \textbf{Pred. CMC (mM)} & \textbf{Meas. CMC (mM)} & \textbf{Source} & \textbf{Temp. ($^\circ$C)} \\
        \hline 
		Outlier 1 & 0.42 & 2.1 & ~\citep{Mukerjee1971} & 25\\
		
		Outlier 2 & 0.27 & 1.2 &~\citep{Mukerjee1971} & 25\\
		
		Outlier 3 & 0.82 & 0.2 &~\citep{MyersAugust2020} &  25\\
		
		Outlier 4 & 11.28 & 42.1 &~\citep{Omar1998} & 34.85\\
		\end{tabular}}
	\end{center}
\end{table}

\begin{table}
	\begin{center}
		\caption{Comparison of predicted vs measured CMC values of the 4 outlier for the distinct surfactant test set. The data source is provided.}
		\label{tab:appendix_outliers_distinct_surfactant}
		\resizebox{\linewidth}{!}{%
		\begin{tabular}{l | c c c c}
		& \multicolumn{4}{ c }{\textbf{Distinct surfactant test set}} \\
        &  \textbf{Pred. CMC (mM)} & \textbf{Meas. CMC (mM)} & \textbf{Source} & \textbf{Temp. ($^\circ$C)} \\
        \hline 
		Outlier 1 & 0.15 & 1.5 & ~\citep{Rosen2012} & 25\\
		
		Outlier 2 & 0.19 & 1.6 &~\citep{Shinoda1977} & 25\\
		
		Outlier 3 & 1.17 & 9.8 &~\citep{Mattei2013} &  25\\
		
		Outlier 4 & 2.24 & 0.4 &~\citep{Xu2016} & 20\\

		\end{tabular}}
	\end{center}
\end{table}

\begin{table}
	\centering
	\caption{Comparison for predicted vs measured CMC values (in mM) of the 3 selected sugar-based surfactants present in the distinct surfactant test set. 
		\textsuperscript{$\alpha$} At 37$^\circ$C.}
	\label{tab:appendix_sugar_based_predictions}
    \begin{tabular}{c|c c |c c | c c}
        \textbf{Temp. (\textdegree C)} & \multicolumn{2}{c}{\textbf{Decyl $\beta$-D-maltoside}} & \multicolumn{2}{c}{\textbf{Decyl $\beta$-D-glucoside}} & \multicolumn{2}{c}{\textbf{Octyl $\beta$-D-Thioglucoside}} \\
        \hline
          & \textbf{Pred.} & \textbf{Meas.} & \textbf{Pred.} & \textbf{Meas.} & \textbf{Pred.} & \textbf{Meas.} \\
        10 & & & 1.82 & 2.53 & & \\
        15 & 2.32 & 2.41 & & & & \\
        20 &  & & 1.84 & 2.09 & 21.4& 9.3 \\
        25 & 2.29 & 2.2 & & &21.45 &8.5 \\
        30 & & & & & 21.73 & 7.5 \\
        35 & 2.3 & 2.05 & 1.94 \textsuperscript{$\alpha$} & 2.03\textsuperscript{$\alpha$} & 22.06 & 6.9 \\
        40 & & & & & 22.42 & 6.2 \\
        45 & 2.35 & 2.1 & & & & \\
        50 & & & 2.01 & 2.13 & & \\
        55 & 2.43 & 2.27 & & & & \\
        65 & 2.53 & 2.57 & & & & \\
        \hline
    \end{tabular}    
\end{table}

\end{document}